\documentclass[lettersize,journal]{IEEEtran}
\usepackage{amsmath,amsfonts}
\usepackage{algorithmic}
\usepackage{algorithm}
\usepackage{array}
\usepackage[caption=false,font=normalsize,labelfont=sf,textfont=sf]{subfig}
\usepackage{textcomp}
\usepackage{stfloats}
\usepackage{url}
\usepackage{verbatim}
\usepackage{graphicx}
\usepackage{cite}
\usepackage{multirow}
\usepackage{hyperref}

\usepackage{amsfonts}
\usepackage{epstopdf}
\usepackage{booktabs}
\usepackage{makecell}

\begin{document}

\title{BDHT: Generative AI Enables
\\
Causality Analysis for Mild Cognitive Impairment}
\author{Qiankun Zuo, Ling Chen, Yanyan Shen, Michael Kwok-Po Ng, Baiying Lei, Shuqiang Wang
\thanks{Qiankun Zuo is with the Hubei Key Laboratory of Digital Finance Innovation, the School of Information Engineering, and also the Hubei Internet Finance Information Engineering Technology Research Center, Hubei University of Economics, Wuhan 430205, Hubei China.(Email: qk.zuo@hbue.edu.cn)}
\thanks{Ling Chen is with the School of Information Management, Institute of Big Data and Digital Economy, Hubei University of Economics, Wuhan 430205, Hubei China.(Email: chenling@hbue.edu.cn)}
\thanks{Michael Kwok-Po Ng is with Department of Mathematics, Hong Kong Baptist University, Hong Kong. (Email:michael-ng@hkbu.edu.hk)}
\thanks{Baiying Lei is with the School of Biomedical Engineering, Shenzhen University, Shenzhen, 518060, China.(Email: leiby@szu.edu.cn)}
\thanks{Yanyan Shen and Shuqiang Wang are with the Shenzhen Institutes of Advanced Technology, Chinese Academy of Sciences, Shenzhen, 518055,China. Email:sq.wang@siat.ac.cn}
}

\markboth{}
{Shell \MakeLowercase{\textit{et al.}}: A Sample Article Using IEEEtran.cls for IEEE Journals}


\maketitle

\begin{abstract}
Effective connectivity estimation plays a crucial role in understanding the interactions and information flow between different brain regions. However, the functional time series used for estimating effective connectivity is derived from certain software, which may lead to large computing errors because of different parameter settings and degrade the ability to model complex causal relationships between brain regions. In this paper, a brain diffuser with hierarchical transformer (BDHT) is proposed to estimate effective connectivity for mild cognitive impairment (MCI) analysis.
To our best knowledge, the proposed brain diffuser is the first generative model to apply diffusion models to the application of generating and analyzing multimodal brain networks.
Specifically, the BDHT leverages structural connectivity to guide the reverse processes in an efficient way. It makes the denoising process more reliable and guarantees effective connectivity estimation accuracy. To improve denoising quality, the hierarchical denoising transformer is designed to learn multi-scale features in topological space. By stacking the multi-head attention and graph convolutional network, the graph convolutional transformer (GraphConformer) module is devised to enhance structure-function complementarity and improve the ability in noise estimation. Experimental evaluations of the denoising diffusion model demonstrate its effectiveness in estimating effective connectivity. The proposed model achieves superior performance in terms of accuracy and robustness compared to existing approaches. Moreover, the proposed model can identify altered directional connections and provide a comprehensive understanding of parthenogenesis for MCI treatment.
\end{abstract}


\def\abstractname{Note to Practitioners}
\begin{abstract}
Diagnosing MCI allows for timely intervention and treatment measures to potentially slow down or even halt further cognitive decline. Exploring causal relations between brain regions enables a better understanding of pathogenic mechanisms and the development of effective biomarkers for MCI diagnosis. The current practice heavily relies on the software to analyze MCI causality, leading to large computing errors and degrading MCI analysis performance because of different parameter settings. This work aims to provide a unified framework for the estimation of brain effective connectivity using generative artificial intelligence. Due to their ability to generate high-quality samples, diffusion models have demonstrated remarkable performance in cross-modal medical image synthesis through iterative denoising processes. Our model provides a new insight into how to transform four-dimensional functional magnetic resonance imaging into effective connectivity without relying on software toolkits. The proposed model achieves good disease prediction performance and identifies altered directional connections that may be potential biomarkers for MCI treatment. Our work enables practitioners to develop deep learning model-based medical tools to assist clinicians with disease diagnosis and pathological analysis in an efficient way. Our work can also extend to the intelligently assisted diagnosis of other neurological diseases.
\end{abstract}


\section{Introduction}
\label{s1}

\IEEEPARstart{M}{ild} cognitive impairment (MCI) is an early stage of a neurodegenerative disorder that primarily affects the brain, leading to progressive cognitive decline\cite{gupta2016traumatic,lei2022predicting}. It typically begins with subtle memory loss and difficulty performing daily tasks and eventually leads to cognitive degradation. Understanding the underlying mechanisms and processes involved in MCI is crucial for developing effective diagnostic tools and potential treatments\cite{myszczynska2020applications}. Effective connectivity (EC) refers to the causal influence that one brain region exerts over another, representing the flow of information and communication within the brain\cite{deshpande2011instantaneous,park2018dynamic}. Studying EC in MCI provides valuable insights into how the disease disrupts the normal functioning of neural networks. The EC helps researchers identify the specific brain regions that are affected and how their interactions are altered\cite{zhong2014altered,xia2023structure}. This can aid in identifying biomarkers for MCI detection, tracking MCI progression, and developing targeted therapeutic interventions.

Effective connectivity (EC) represents a directed brain network, typically derived from functional magnetic resonance imaging (fMRI) for MCI analysis. In contrast to methods utilizing undirected brain networks (i.e., functional connectivity, structural connectivity, or multimodal connectivity)\cite{lei2021diagnosis,gong2023generative,zuo2023alzheimer} or imaging-based data\cite{wang2024novel}, EC-based approaches focus on the causal interactions between regions of interest (ROIs). They incorporate additional directional information between ROIs, resulting in enhanced performance in diagnosing MCI.
Estimating EC for brain disease analysis has been a hot topic for a long time\cite{cao2022brain}. Traditional methods have been designed to extract EC-based features for brain disease analysis. For example, the work in \cite{cosio2022diagnosis} applied structural equation modeling (SEM) to construct EC and analyzed the late-life depression. Li \emph{et al.}\cite{li2020large} constructed EC using dynamic causal modeling (DCM) and proved its robustness in detecting altered causal connections for major depressive disorder. Crimi \emph{et al.}\cite{jiang2021analysing} modified the Granger causality models (GCM) and proposed the nonlinear Granger kernel-based method to estimate the EC for deductive reasoning tasks.
Different imaging modalities capture different aspects of brain connectivity features\cite{pan2021characterization, zuo2023prior}, integrating multimodal imaging can enhance complementarity and allow for a more complete evaluation of EC. Scherr \emph{et al.}\cite{scherr2021effective} utilized the metabolic connectivity mapping (MCM) method by combining fMRI and positron emission tomography (PET) to measure EC for brain disorder analysis. Also, Ji \emph{et al.}\cite{ji2019acoec} explored the combination of fMRI and diffusion tensor imaging (DTI) to build the Acoec-fd model for learning EC and achieved promising results in cognitive disease diagnosis. However, these methods often rely on assumptions and simplifications about the underlying neural dynamics, which can limit the ability to capture nonlinear interactions or time-varying dynamics, potentially leading to a biased or incomplete estimation of EC.

The deep learning-based methods have achieved good performance in medical image analysis because of their great ability in high-level feature learning\cite{wu2023multi,wang2020ensemble,hu2020medical,hu20233}. Deep learning techniques perform well at modeling causal relationships between brain areas, which allows for exploring the pathogenesis of brain diseases.
The study in \cite{yang2023spatial} proposed a spatial-temporal directed acyclic graph convolutional network (ST-DAGCN) to model direct causal connections from fMRI.
The work in \cite{zhang2023amortization} also modeled causal connectivity from fMRI for brain disease analysis.
To improve the accuracy of EC estimation, generative adversarial network (GAN)-based methods are developed to learn complex data distributions for modeling causal relations. The work in\cite{ji2021estimating} explored the combination of recurrent neural networks and GAN and proposed the EC-RGAN model to estimate EC for temporal feature preservation. And Zou \emph{et al.}\cite{zou2022exploring} considered the non-Euclidean topological properties and constructed EC through the spatiotemporal graph convolutional models (STGCM). They captured the spatial-temporal topological relations between brain regions and achieved good performance in simulated and real datasets.
Considering the dynamic characteristics of EC, Ji \emph{et al.}\cite{ji2023dynamic} proposed a GAN-based nonparametric state estimation model to accurately learn the temporal patterns of EC.
Furthermore, Liu \emph{et al.}\cite{liu2024mcan} utilized a GAN-based model to estimate the dynamic EC from multimodal imaging data. They achieved good prediction results and revealed abnormal connectivity-based patterns in the visual categorization task. Similar work on constructing EC using multiple modalities is founded in \cite{10097497}.
Besides, Ji \emph{et al.}\cite{ji2024metacae} developed a meta-knowledge transfer-based causal autoencoder to estimate EC to solve the problem of limited fMRI data.
However, the current way of constructing EC from fMRI is to first preprocess the fMRI with the software toolbox to obtain the empirical ROI-based time series (clean sample) and then build a deep learning model to construct EC.
The main drawbacks of current methods lie in that they heavily rely on the empirical ROI-based time series to construct EC, ignoring the manual parameter setting differences during preprocessing procedures, thus introducing large errors in EC estimation.

The generative models\cite{zhou2023scgan,li2023dls,wang2020diabetic} are suitable to learn a mapping from imaging data to the effective connectivity feature for two main reasons: (1) The lack of medical imaging data is very common and may lead to overfitting problems in feature extraction. The generative models can increase the diversity of training data by generating new samples, thereby improving the generalization ability and performance of the models. (2) The distribution of brain connectivity features is very complex and hard to learn. The generative models can empower the modeling of implicit data distribution and capture the complex directional relationships within the brain regions.
The straightforward way to generate ROI-based time series and construct EC from fMRI is to build a GAN-based model with one generator and one discriminator. When the discriminator cannot distinguish the data distribution, the optimized generator can map the fMRI into EC in an end-to-end manner. However, the issue of instability and mode collapse during adversarial training still has to be fully resolved \cite{hu2020brain,wang2022brain}, which limits its ability to be used in this situation. Recently, denoising diffusion probabilistic models (DDPM)\cite{ho2020denoising,nichol2021improved} have gained significant attention in the field of deep learning and have shown promising results in various domains, including image generation and audio generation. The fundamental idea behind DDPM is to model the data distribution as a diffusion process, where noise is gradually transformed into a clean sample through a sequence of intermediate denoising steps. It has shown notable benefits in generating high-quality samples with improved mode coverage and stability compared to other generative models like GANs.
Furthermore, as the DTI provides the basic physical connection between brain regions, the DTI is coupled with the fMRI in cognitive activities\cite{park2013structural}. Combining both of them can offer a more solid foundation for time series generation and EC construction.

Inspired by the above observations, the brain diffuser with hierarchical transformer (BDHT) is proposed to estimate EC from fMRI in an end-to-end framework.
It is a modified DDPM that leverages the conditional fMRI and DTI to guide the denoising processes in an efficient way. The transformer-based network is designed to capture multi-scale topological features and enhance structure-function complementarity for better noise removal and EC estimation.
The proposed BDHT is a unified AIGC framework to transform four-dimensional fMRI into effective connectivity without relying on software toolkits.
The BDHT demonstrates robust MCI prediction performance and identifies altered directional connections that may be potential biomarkers for MCI treatment.
The main contributions are summarized as follows:
\begin{itemize}
	\item The proposed BDHT is a unifying generative framework that leverages the diffusion-based model to estimate effective connectivity with several successive denoising steps. The proposed model provides a new insight into how to transform four-dimensional fMRI into effective connectivity using generative artificial intelligence, which is efficient and reliable in effective connectivity estimation for MCI analysis.
	
	\item The hierarchical denoising transformer is designed to learn multi-scale features for noise removal during the reverse denoising process. Moreover, the structural connectivity guides the model to capture topologically directional features by modeling relationships between distant brain regions. These make the denoising process interpretable and guarantee that the effective connectivity estimation is more accurate.
	
	\item By stacking the multi-head attention and graph convolutional network, the graph convolutional transformer (GraphConformer) block is devised to concentrate on both global and adjacent connectivity information, which enhances structure-function complementarity and improves its ability to estimate noise.
	
\end{itemize}

The organization of this work is structured as follows: The main architecture of the proposed BDHT model is introduced in Section ~\ref{s3}. In Section ~\ref{s4}, we implement the proposed model on the public datasets and analyze the results. Section ~\ref{s5} shows the reliability of our results, and Section ~\ref{s6} summarizes the main remarks of this work.

\begin{figure}[htbp]
	\centering
	\includegraphics[width=\columnwidth]{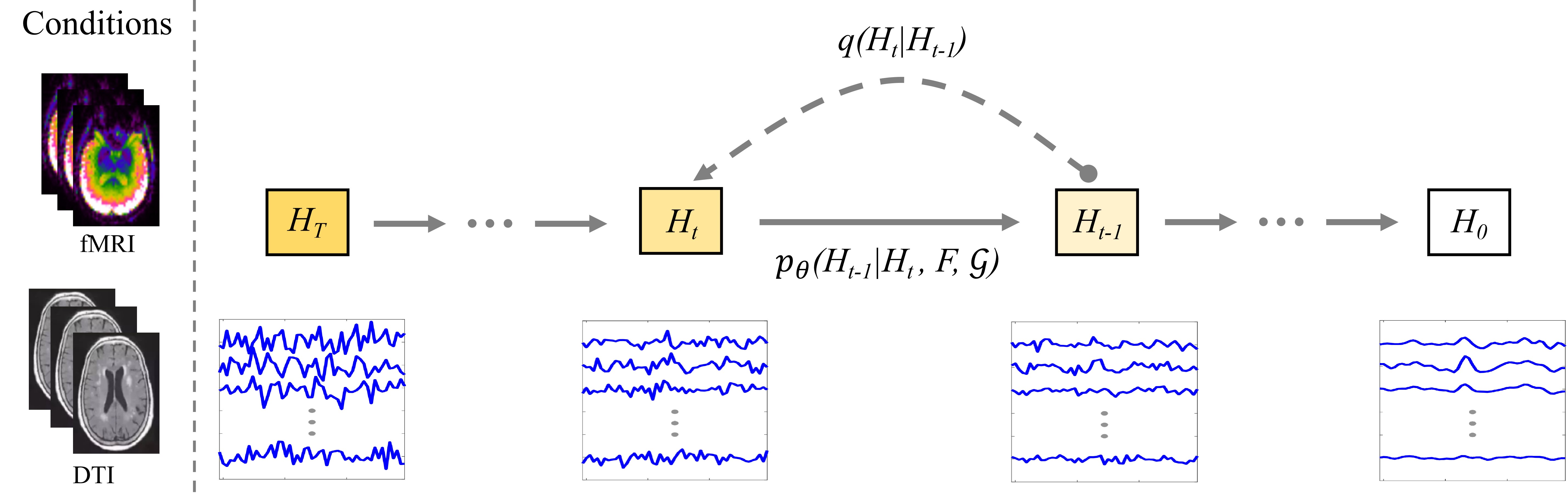}
	\caption{The schematic diagram of the proposed BDHT model, including the diffusive process (dashed arrows) and the denoising process (solid arrows). The two conditions are the fMRI and DTI. \label{fig0}}
\end{figure}

\begin{figure*}[htbp]
	\centering
	\includegraphics[width=\textwidth]{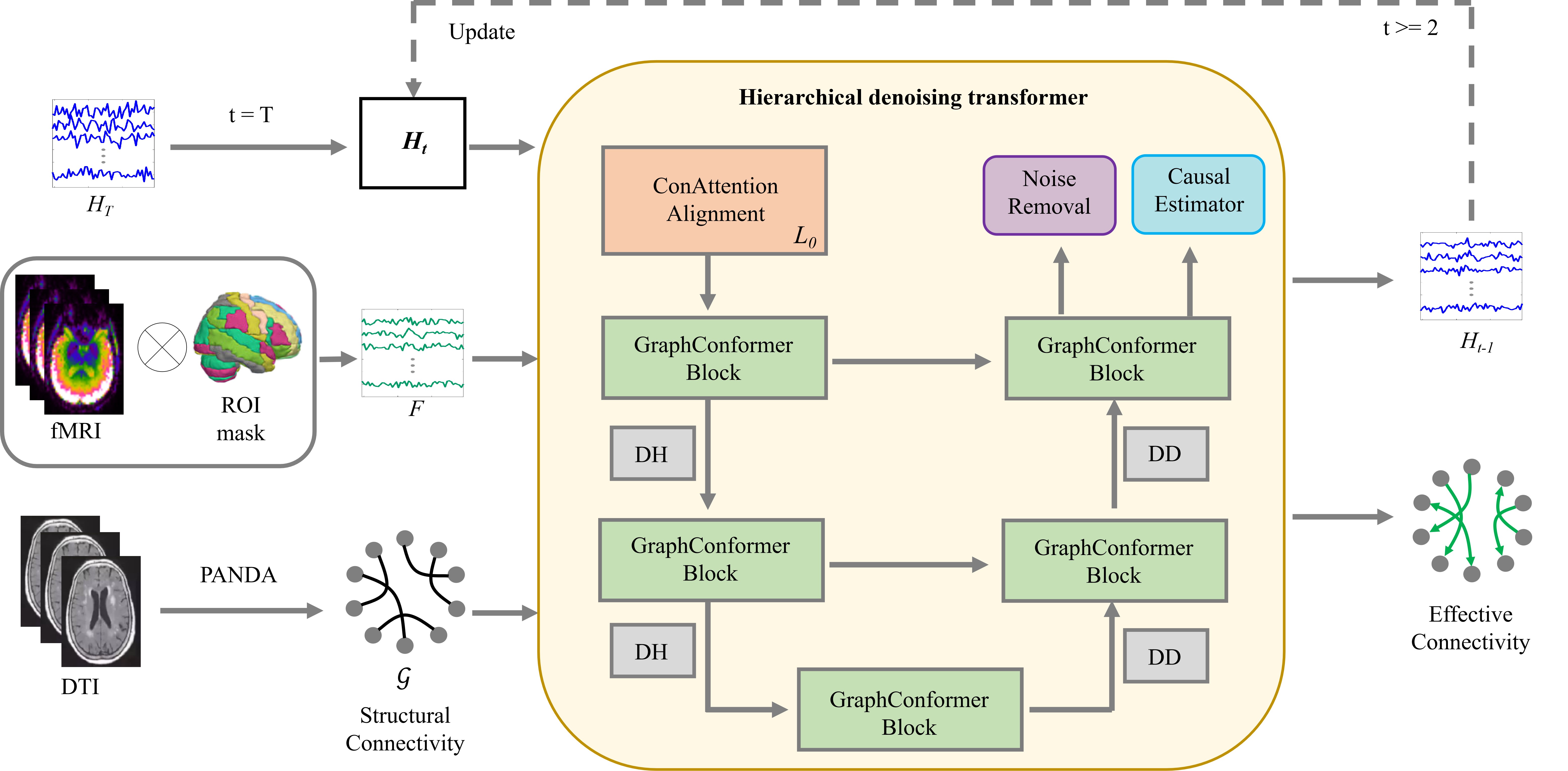}
	\caption{ The detailed network structure of one denoising step in the proposed BDHT model. In the diffusive process, the empirical sample $\boldsymbol{H}_0$ is gradually transformed into the Gaussian sample $\boldsymbol{H}_T$ by adding noise at each successive step. In the denoising process, the rough sample $\boldsymbol{F}$ and structural connectivity $\mathcal{G}$ are input as conditions to guide the hierarchical denoising transformer to generate the clean sample and effective connectivity.\label{fig1}}
\end{figure*}

\section{Method}
\label{s3}

The proposed BDHT aims to transform fMRI into effective connectivity for MCI analysis.
As shown in Fig.~\ref{fig0}, the proposed BDHT is a modified DDPM by adding two conditions (fMRI and DTI). The right side of the vertical dotted line in Fig.~\ref{fig0} is the conventional DDPM.
The diffusive process of BDHT is the same as the DDPM: the empirical sample $\boldsymbol{H}_0$ is transformed to the Gaussian sample $\boldsymbol{H}_T$ by gradually adding Gaussian noise through a series of successive steps (left arrows in Fig.~\ref{fig0}).
In the denoising process (right arrows in Fig.~\ref{fig0}), the fMRI and DTI are input as conditions to guide the hierarchical denoising transformer to generate the clean sample (also called clean ROI-based time series) and effective connectivity.
Fig.~\ref{fig1} shows one detailed computing step during the denoising process of the BDHT model.
The fMRI is preprocessed to obtain the empirical ROI-based time series (an empirical sample generated by the software toolbox) $\boldsymbol{H}_0$ and the rough ROI-based time series (rough sample derived by the non-parameter operations) $\boldsymbol{F}$.
The structural connectivity $\mathcal{G}$ derived from DTI and the empirical sample guide the hierarchical denoising transformer to denoise the sample $\boldsymbol{H}_t$ at the $t$-th step into the sample $\boldsymbol{H}_{t-1}$ at step $t-1$.
In the following, we first describe the basic ideas of DDPM, then present the framework of the proposed BDHT, and finally introduce the important network structure of the hierarchical denoising transformer in the BDHT.


\subsection{Basic ideas of DDPM}
DDPM is a generative modeling framework that aims to learn a Markov chain by modeling the underlying probability distribution between Gaussian noise and clean data. Assuming the clean data distribution $q(\boldsymbol{H}_{0})$, DDPM can learn a distribution $p_\theta(\boldsymbol{H}_{0})$ from the Gaussian distribution to approach $q(\boldsymbol{H}_{0})$. The distribution modeling can be divided into two parts: the diffusive process and the denosing process.

In the diffusive direction, the clean sample $\boldsymbol{H}_{0} \sim q(\boldsymbol{H}_{0})$ is accumulatively added by the Gaussian noise to obtain the Gaussian sample $\boldsymbol{H}_{T} \sim \mathcal{N}(\mathbf{0},\boldsymbol{I})$ with sufficient $T(T \gg 1)$ steps. At each step, a weight is applied to the Gaussian noise to form a Markov chain. The applied weights follow a variance schedule: $\beta_{1}, \beta_{2},..., \beta_{T}$. For simplification, the intermediate noisy sample $\boldsymbol{H}_{t}$ is computed from the sample at step $t-1$ with the following formulas:
\begin{equation}\label{eq1}
	\boldsymbol{H}_{t}=\sqrt{1-\beta_{t}} \boldsymbol{H}_{t-1}+\sqrt{\beta_{t}} \boldsymbol{\epsilon}, \quad \boldsymbol{\epsilon} \sim \mathcal{N}(\mathbf{0}, \boldsymbol{I})
\end{equation}
\begin{equation}
		q\left(\boldsymbol{H}_{t} \mid \boldsymbol{H}_{t-1}\right)=\mathcal{N}\left(\boldsymbol{H}_{t} ; \sqrt{1-\beta_{t}} \boldsymbol{H}_{t-1}, \beta_{t} \boldsymbol{I}\right)
\end{equation}
where $t$ ranges from $1$ to $T$, $\quad \boldsymbol{\epsilon}$ means the added Gaussian noise, $\mathcal{N}$ represents the Gaussian distribution, and $\boldsymbol{I}$ is a multivariate covariance identity matrix. $\beta_{t} \in (0,1)$ increases with the growth of time steps. After $T$ steps, the Gaussian sample $\boldsymbol{H}_{T}$ can be expressed by:
\begin{equation}\label{eq3}
\boldsymbol{H}_{T}=\sqrt{\hat{\alpha}_{t}} \boldsymbol{H}_{0}+\sqrt{1-\hat{\alpha}_{t}} \boldsymbol{\epsilon}_{t}
\end{equation}
here, $\alpha_{t}=1-\beta_{t}, \hat{\alpha}_{t}=\prod_{j=1}^{t} \alpha_{j}$, $\boldsymbol{\epsilon}_{t}$ are sampled from the Gaussian distribution with the same size of $\boldsymbol{H}_{t}$. A large $T$ can ensure the noisy sample $\boldsymbol{H}_{t}$ approximates a standard normal distribution.

During the denoising direction, the DDPM learns a Markov chain by gradual denoising the Gaussian sample $\boldsymbol{H}_{T}$ to the clean sample $\boldsymbol{H}_{0}$. The assumptions of large $T$ and small $\beta_{t}$ guarantee adjacent samples follow the Gaussian distribution. The denoising probability from $\boldsymbol{H}_{t}$ to $\boldsymbol{H}_{t-1}$ can be defined as:
\begin{equation}\label{eq4}
p_{\theta}\left(\boldsymbol{H}_{t-1} \mid \boldsymbol{H}_{t}\right)=\mathcal{N}\left(\boldsymbol{H}_{t-1} ; \boldsymbol{\mu}_{\theta}\left(\boldsymbol{H}_{t}, t\right), \boldsymbol{\Sigma}_{\theta}\left(\boldsymbol{H}_{t}, t\right)\right)
\end{equation}
where the $\boldsymbol{\mu}_{\theta}(\boldsymbol{H}_{t}, t)$ and $\boldsymbol{\Sigma}_{\theta}(\boldsymbol{H}_{t}, t)$ are the distribution parameters that can be learned by a designed neural network. To optimize the parameterized network, the Kullback-Leibler (KL) divergence is introduced to minimize the distance between the estimated distribution $p_{\theta}(\boldsymbol{H}_{t-1} \mid \boldsymbol{H}_{t})$ and the posterior probability $q(\boldsymbol{H}_{t-1}; \boldsymbol{H}_{t}, \boldsymbol{H}_{0})$. After simplifying the KL formula, the Gaussian distribution parameter can be derived:
\begin{equation}
\boldsymbol{\mu}_{\theta}\left(\boldsymbol{H}_{t}, t\right)=\frac{1}{\sqrt{\alpha_{t}}}\left(\boldsymbol{H}_{t}-\frac{\beta_{t}}{\sqrt{1-\hat{\alpha}_{t}}} \boldsymbol{\epsilon}_{\theta}\left(\boldsymbol{H}_{t}, t\right)\right)
\end{equation}
\begin{equation}
\boldsymbol{\Sigma}_{\theta}(\boldsymbol{H}_{t}, t) = \frac{\left(1-\hat{\alpha}_{t-1}\right)}{\left(1-\hat{\alpha}_{t}\right)} \beta_{t} \boldsymbol{I}
\end{equation}
The learnable function $\boldsymbol{\epsilon}_{\theta}$ estimates the added noise in the diffusive process, which should be removed at the denoising process. The estimated noise can be computed by the following function:
\begin{equation}
\operatorname*{min}_{\theta}\mathbb{E}_{\boldsymbol{H}_{0},\boldsymbol{\epsilon}_{t},t}\left\|\boldsymbol{\epsilon_t}-\boldsymbol{\epsilon}_{\theta}(\sqrt{\hat{\alpha}_{t}} \boldsymbol{H}_{0}+\sqrt{1-\hat{\alpha}_{t}} \boldsymbol{\epsilon}_{t},t)\right\|^{2}
\end{equation}
Once the network is optimized, the Gaussian noise can be transmitted to a clean sample by removing the noise step by step.

\begin{figure}[htbp]
	\centering
	\includegraphics[width=0.8\columnwidth]{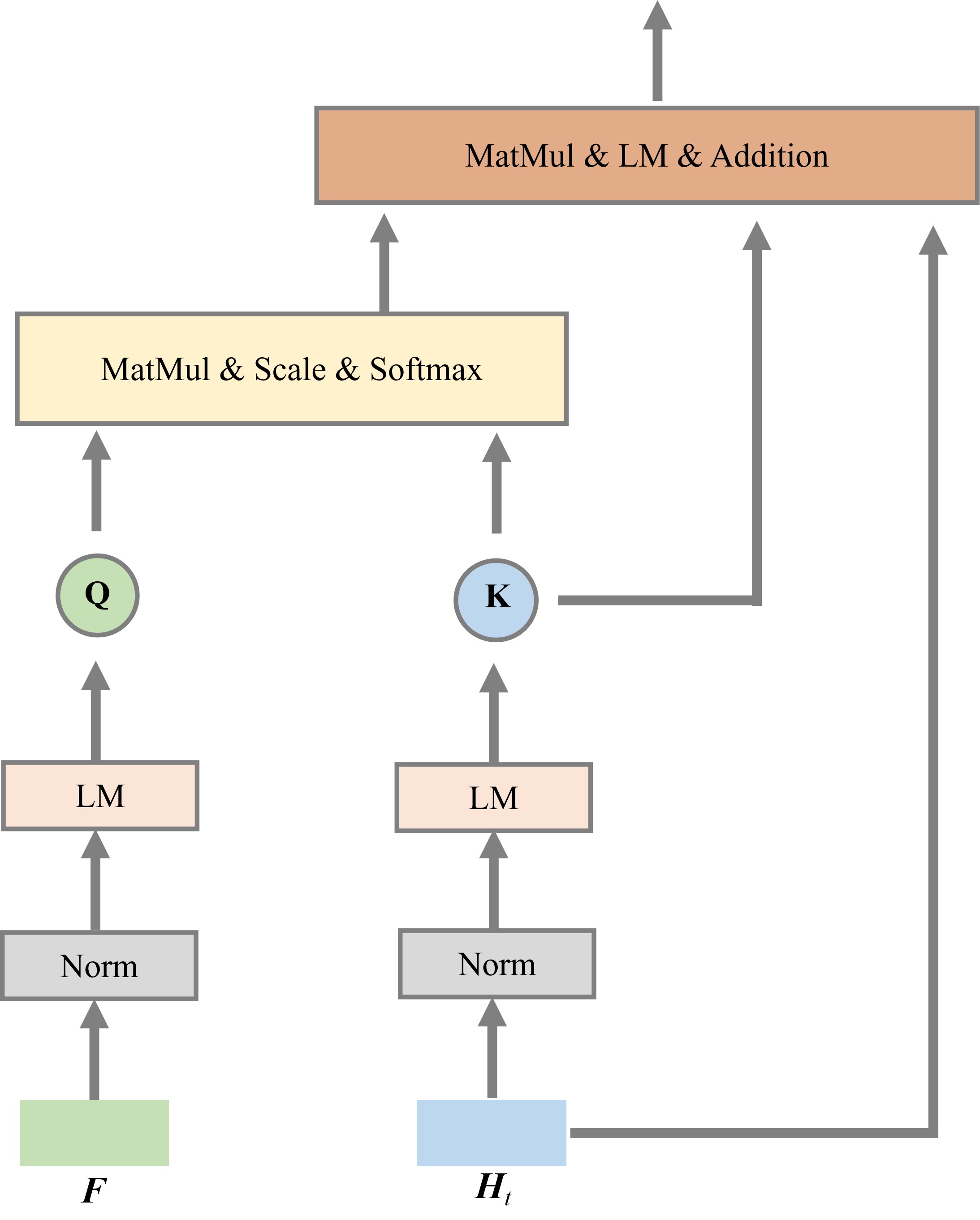}
	\caption{The structure of the ConAttention Alignment. The inputs are the rough sample $\boldsymbol{F}$ and intermediate noisy sample $\boldsymbol{H}_t$, and the output is the fused feature.\label{fig2}}
\end{figure}

\begin{figure*}[t]
	\centering
	\includegraphics[width=\textwidth]{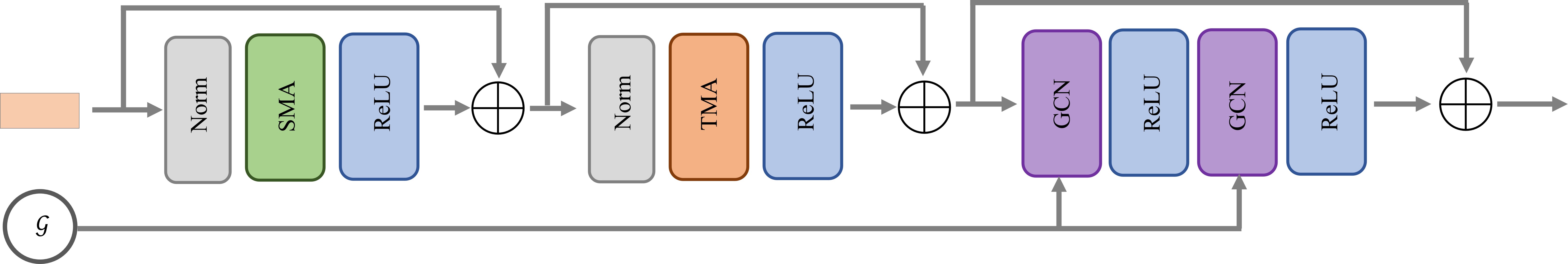}
	\caption{The computing structure of the GraphConFormer block. The input is the fused feature, and the output updates the input and shares the same dimension. The $\mathcal{G}$ guides the topological feature learning in the GCN layers.\label{fig3}}
\end{figure*}

\subsection{Brain diffuser with hierarchical transformer}
The conventional DDPM is to generate clean samples from Gaussian noise without any conditions, which does not meet our task of converting fMRI into effective connectivity. Therefore, we modify the DDPM and propose the brain functional diffuser by adding fMRI as a condition in the denoising direction. Besides, considering the topological properties can provide rich information about the estimated noise in the denoising process, we add another condition, which is the structural connectivity $\mathcal{G}$.

In our model, we first transformed the fMRI into rough ROI-based time series (rough sample), then combined it with the structural connectivity $\mathcal{G}$ to guide the DDPM for noise removal and clean sample generation. The rough sample $\boldsymbol F$ can be derived by the following steps: (1) at each time point of fMRI, the 3D MRI volume was warped into several ROIs by multiplying the standard 3D anatomical atlas ($all.nii$); (2) all the pixels in each ROI were summed to obtain one value; (3) we concatenated the obtained values of all the time points for each ROI; (4) we removed the first 10 time points and derived the rough ROI-based time series $\boldsymbol{F}$ with the dimension size $90 \times 187$.
As shown in Fig.~\ref{fig0}, the diffusive process of the brain functional diffuser is the same as Eq.(~\ref{eq1}), while the denoising process can be expressed as follows:
\begin{equation}\label{eq8}
\begin{aligned}
	& p_{\theta}\left(\boldsymbol{H}_{t-1} \mid \boldsymbol{H}_{t}, \boldsymbol{F},\mathcal{G}\right)= \\
     &\mathcal{N}\left(\boldsymbol{H}_{t-1} ; \boldsymbol{\mu}_{\theta}\left(\boldsymbol{H}_{t}, t \mid \boldsymbol{F},\mathcal{G}\right), \boldsymbol{\Sigma}_{\theta}\left(\boldsymbol{H}_{t}, t \mid \boldsymbol{F},\mathcal{G}\right)\right)
\end{aligned}
\end{equation}
\begin{equation}
p_{\theta}\left(\boldsymbol{H}_{0: T} \mid \boldsymbol{F},\mathcal{G}\right)=p\left(\boldsymbol{H}_{T}\right) \prod_{t=1}^{T} p_{\theta}\left(\boldsymbol{H}_{t-1} \mid \boldsymbol{H}_{t}, \boldsymbol{F},\mathcal{G}\right)
\end{equation}

We denote the causal estimator as $\boldsymbol{\phi}_{\theta}$ consisting of two multilayer perceptron (MLP) layers.  The structure of the transformer-based network is denoted as $\boldsymbol{\epsilon}_{\theta}$. The predicted noisy sample $\boldsymbol{H}_{t-1}'$, the reconstructed noisy sample $\hat {\boldsymbol{H}} _{t-1}$ and the effective connectivity $\mathbf{E}$ are defined as:
\begin{equation}\label{eq10}
	\begin{aligned}
		\boldsymbol{H}_{t-1}'= & \sqrt{\hat{\alpha}_{t-1}}\left(\frac{\boldsymbol{H}_{t}-\sqrt{1-\hat{\alpha}_{t}} \cdot \boldsymbol{\epsilon}_{\theta}\left(\boldsymbol{H}_{t}, t \mid \boldsymbol{F},\mathcal{G}\right)}{\sqrt{\hat{\alpha}_{t}}}\right) \\
		& +\sqrt{1-\hat{\alpha}_{t-1}} \cdot \boldsymbol{\epsilon}_{\theta}\left(\boldsymbol{H}_{t}, t \mid \boldsymbol{F},\mathcal{G}\right)
	\end{aligned}
\end{equation}
\begin{equation}\label{eq11}
	\hat {\boldsymbol{H}} _{t-1} = \boldsymbol{\phi}_{\theta} (\boldsymbol{H} _{t}, \boldsymbol{\epsilon}_{\theta}, t)
\end{equation}

During the optmization, we sample $(\boldsymbol{H}_{0}, \boldsymbol{F}, \mathcal{G}) \sim q(\boldsymbol{H}_{0}, \boldsymbol{F}, \mathcal{G})$ from the truth data distribution. The denoising process can be trained by the following objective function:
\begin{equation}
\begin{aligned}
	\mathcal{L}(\theta)=  &\mathbb{E}_{\boldsymbol{H}_{t}, \boldsymbol{\epsilon}_{t}, t}\left\|\boldsymbol{\epsilon_t}-\boldsymbol{\epsilon}_{\theta}
	\left(\boldsymbol{H}_{t}, t \mid \boldsymbol{F},\mathcal{G}\right)
	\right\|^{2} \\
	& + \mathbb{E}_{\boldsymbol{H}_{t-1}, \boldsymbol{\epsilon}_{\theta}, t} \left\| \boldsymbol{H}_{t-1} - \mathbf{E}^{transpose} \cdot \hat {\boldsymbol{H}} _{t-1}  \right\|^{2}  \\
    & + \frac{\omega}{N*N} \left\| \mathbf{E}\right\|
\end{aligned}
\end{equation}
where, the $\omega$ is a sparsity constraint parameter to overcome the overfitting problem.
Thousands of sampling steps lead to slow denoising computation; we reduce the sampling steps by deviding the whole $T$ steps into several sub-steps. Assuming the new total sampling steps are $T_n$ and the sampling interval is $m$, we can define the new sampling step formula: $j = floor(i/m)+1, i \in {1,2,...,T}, j \in {1,2,...,T_n}$.

\subsection{Hierarchical denoising transformer}
The inputs of the denoising network are the intermediate noisy sample $\boldsymbol{H}_t \in \mathbb{R}^{N \times q}$, the rough (conditional) sample $\boldsymbol{F} \in \mathbb{R}^{N \times q}$, and the structural connectivity $\mathcal{G} \in \mathbb{R}^{N \times N}$. Note that $N$ is the number of ROIs, and $q$ is the dimension of each sample. The $\boldsymbol{F}$ is obtained by pixel multiplication between the fMRI and the anatomical ROI mask. The $\mathcal{G}$ is estimated from diffusion tensor imaging (DTI) by the GRETNA software. During the denoising computation, there are three main modules, including the conditional attention alignment (ConAttention alignment), the graph convolutional transformer (GraphConFormer), the noise removal and the causal estimator. After the denoising is complete, the classifier is adopted to evaluate the discriminability of the generated effective connectivities.

\subsubsection{Conditional attention alignment}
The current approach of fusing the conditional sample and the intermediate noisy sample is to concatenate them \cite{guo2023shadowdiffusion}, which leads to high dimensions in our study. As shown in Fig.~\ref{fig2}, we design one-layer linear mapping (LM) to obtain the query ($\boldsymbol Q$) and key ($\boldsymbol K$). An attention layer is applied to the $\boldsymbol Q$ and $\boldsymbol K$ to compute the fused feature $\boldsymbol{H}_t^{ConAttention}$. The fused ROI-based feature can be well aligned between the conditional sample $\boldsymbol{F}$ and the intermediate noisy sample $\boldsymbol{H}_t$, enhancing the signal and suppressing the noise. The calculation process is as follows:
\begin{equation}
	\boldsymbol{Q} = LM_1(Norm(\boldsymbol{F})), \boldsymbol{K} = LM_2(Norm(\boldsymbol{H}_t))
\end{equation}
\begin{equation}
	\mathcal{X}=\operatorname{softmax}\left(\frac{\boldsymbol{Q} \boldsymbol{K}^T}{\sqrt{q}}\right) \boldsymbol{K}
\end{equation}
\begin{equation}
	\boldsymbol{H}_t^{ConAttention} = LM_3(\mathcal{X}) + \boldsymbol{H}_t
\end{equation}
After $L_0$ ConAttention blocks, the fused feature $\boldsymbol{H}_t^{ConAttention}$ has the same size with the $\boldsymbol{H}_t  \in \mathbb{R}^{N \times q}$.

\subsubsection{Graph convolutional transformer}
To learn multi-scale features in topological space for noise removal, we designed a U-shaped structure consisting of GraphConFormer, dimension halving (DH), and dimension doubling (DD). As shown in Fig.~\ref{fig3}, the GraphConFormer block stacks multi-head attention networks and graph convolutional networks (GCN) \cite{kipf2016semi} for extracting both global and adjacent connectivity information.
To be specific, the fused feature $\boldsymbol{H}_t^{ConAttention}$ first passes the spatial multi-head attention (SMA), then passes the temporal multi-head attention (TMA). The spatiotemporal feature can be greatly enhanced. Finally, the structural connectivity $\mathcal{G}$ guides the model to learn topological features by using two graph convolutional networks. The detailed computation steps are given as follows:
\begin{equation}
	\Gamma _1 = ReLU(SMA(Norm(\Gamma _0))) + \Gamma _0
\end{equation}
\begin{equation}
	\Gamma _2 = ReLU(TMA(Norm(\Gamma _1))) + \Gamma _1
\end{equation}
\begin{equation}
	\boldsymbol{H}_t^{GraphConFormer} = ReLU(GCN(ReLU(GCN(\Gamma _2, \mathcal{G})))) + \Gamma _2
\end{equation}
here, the input $\Gamma _0$ represents the $\boldsymbol{H}_t^{ConAttention}$ in the first layer, and the variables $\Gamma _1$ and $\Gamma _2$ represent the operations of SMA and TMA, respectively. For SMA, the sequence is the ROI temporal feature, and the heads are set as $h_s$; for TMA, the sequence is the features perpendicular to the temporal direction, and the heads are set as $h_t$. The output $\boldsymbol{H}_t^{GraphConFormer}$ has the same size as $\Gamma _0$.
The input and output of each graph conformer block are the same: the first and fifth blocks have the size $N \times q$, the second and fourth blocks have the size $N \times (q/2)$, and the third block has the size $N \times (q/4)$.

The DH module is a GCN layer, which reduces the feature dimension. For the first DH block, the input size is $N \times q$, and the output size is $N \times q/2$. The DD module is also a GCN layer for dimension increase. For the first DD block, the input dimension is $N \times q/4$, and the output dimension is $N \times q/2$. The arrow between GraphConFormer blocks means adding operations.

\begin{algorithm}[h]
	\caption{Training of the BDHT model}
	\label{Algorithm1}
	\begin{algorithmic}[1]
		\REQUIRE
		$\boldsymbol{H}_0$: empirical sample\newline
		\hspace*{1.2em} $\boldsymbol{F}$: conditional sample\newline
		\hspace*{1.2em} $T$: number of diffusion steps\newline
		\hspace*{1.2em} $\mathcal{G}$: structural connectivity
		\ENSURE
		Noise removal function $\boldsymbol{\epsilon}_{\theta}$ and causal estimator $\boldsymbol{\phi}_{\theta}$
		\STATE initial variance schedle {$\beta_{i},i=1,2,...,T$};  $\omega=2.5$
		\REPEAT
		\STATE $t \sim Uniform({1,2,...,T})$,$\boldsymbol{H}_0 \sim q(\boldsymbol{H}_0)$
		\STATE Sample $\boldsymbol{\epsilon}_{t} \sim \mathcal{N}(\mathbf{0}, \boldsymbol{I})$
		\STATE Calculate the diffusive sample $\boldsymbol{H}_t$ using Eq.(~\ref{eq3})
		\STATE Take gradient descent step on $\nabla_{\theta} \mathcal{L}(\theta)$
		\UNTIL converged
	\end{algorithmic}
\end{algorithm}

\subsubsection{Noise removal and Causal estimator}
The aforementioned GraphConFormer outputs the estimated noise $\boldsymbol{\epsilon}_{\theta}\left(\boldsymbol{H}_{t}, t \mid \boldsymbol{F},\mathcal{G}\right)$, we utilize the Eq.(~\ref{eq10}) to calculate the denoised sample $\boldsymbol{H}_{t-1}'$ by the noise removal module. The output at $t=0$ clean sample is defined as $\boldsymbol{H}_0' \in \mathbb{R}^{N \times q}$, which is similar to $\boldsymbol{H}_0$.

In the causal estimator, we explore the causal relations between any pair of brain regions. From $t-$th step to $(t-1)$-th step, we first concatenate the estimated noise $\boldsymbol{\epsilon}_{\theta}$ and the noisy sample $\boldsymbol{H}_t$; then send them to two layers of multi-layer perceptron (MLP) and obtain the reconstructed noisy sample $\hat {\boldsymbol{H}} _{t-1} \in \mathbb{R}^{N \times q}$. Finally, we utilize the following formula to model the signal flow from one brain region to another.
Assuming $\boldsymbol{x}_i$ represents the $i$-th row vector of the $\hat {\boldsymbol{H}} _{t-1}$, all brain regions exerting effects on the $i$-th ROI can be expressed by:
\begin{equation}
\boldsymbol{x}_{i}^{\prime}=\sum_{j=1}^{N} \mathbf{E}_{j i} \boldsymbol{x}_{j}+\boldsymbol{\varepsilon}_{i}
\end{equation}
where $\boldsymbol{\varepsilon}_{i}$ represents the error when reconstructing the target signal. The $\boldsymbol{x}_{i}^{\prime}$ is the $i$-th row vector of the intermediate noisy sample $\boldsymbol{H}_{t-1}$. The calculated matrix $\mathbf{E}$ is asymmetric with a size of $N \times N$. The diagonal elements of $\mathbf{E}$ are zeros because we don't consider the causal effects of the brain region itself. The estimated effective connectivity matrices are sent into the BrainNetCNN classifier \cite{kawahara2017brainnetcnn} for classification evaluation. The BrainNetCNN classifier takes into account the topological locality of brain connectivities by designing CNN-based networks.

The pseudo-code of the training and sampling processes is displayed in Algorithm~\ref{Algorithm1} and Algorithm~\ref{Algorithm2}.

\begin{algorithm}[h]
	\caption{Sampling of the BDHT model}
	\label{Algorithm2}
	\begin{algorithmic}[1]
		\REQUIRE
		$\boldsymbol{F}$: conditional sample\newline
		\hspace*{1.2em} $T$: number of diffusion steps\newline
		\hspace*{1.2em} $\mathcal{G}$: structural connectivity\newline
		\hspace*{1.2em} $\boldsymbol{\epsilon}_{\theta}$: noise estimation function\newline
		\hspace*{1.2em} $\boldsymbol{\phi}_{\theta}$: causal estimation function
		\ENSURE
		Clean sample $\boldsymbol{H}_0'$ and effective connectivity $\mathbf{E}$
		\STATE Sample $\boldsymbol{\epsilon}_{t} \sim \mathcal{N}(\mathbf{0}, \boldsymbol{I})$
		\FOR {$t=T$ to $1$}
		\STATE Calculate the intermediate sample $\boldsymbol{H}_{t-1}'$ using Eq.(~\ref{eq10}) \newline
		\hspace*{1.2em} by taking $\boldsymbol{F}$ and $\mathcal{G}$ as conditions
		\STATE Update the effective connectivity $\mathbf{E}$ using Eq.(~\ref{eq11})
		\ENDFOR
		\STATE Return $\boldsymbol{H}_{0}'$ and $\mathbf{E}$
	\end{algorithmic}
\end{algorithm}

\section{Experiments}
\label{s4}

\subsection{Datasets}
Our study was conducted on the publicly available Alzheimer's Disease Neuroimaging Initiative (ADNI) dataset. We selected the same number of subjects for each category, including significant memory concern (SMC), early mild cognitive impairment (EMCI), late mild cognitive impairment (LMCI), and normal control (NC). The total number of subjects is 240, scanned with both functional magnetic resonance imaging (fMRI) and diffusion tensor imaging (DTI). The male-to-female ratio is nearly balanced.

The fMRI scans were performed by Siemens using the following parameters: a repetition time (TR) of 3.0 s and an echo-planar imaging (EPI) sequence consisting of 197 volumes. The preprocessing approach for the fMRI data requires the anatomical automatic labeling (AAL90) atlas \cite{tzourio2002automated} to compute region of interest (ROI)-based time series. This approach involves using the GRETNA software to obtain the functional time series, which serves as the empirical sample $\boldsymbol{H}_{0}$ in our proposed model. The specific computation steps using GRETNA \cite{wang2015gretna} are described in the work\cite{zuo2022constructing}.
The DTI scanning directions range from 6 to 126. To preprocess the DTI, we utilized the PANDA toolbox \cite{cui2013panda} to first determine brain fractional anisotropy and then output a structural connectivity matrix based on the deterministic fiber tracking method. The subsequent connectivity matrix has a dimension of $90 \times 90$.

\begin{figure}[h]
	\centering
	\includegraphics[width=0.98\columnwidth]{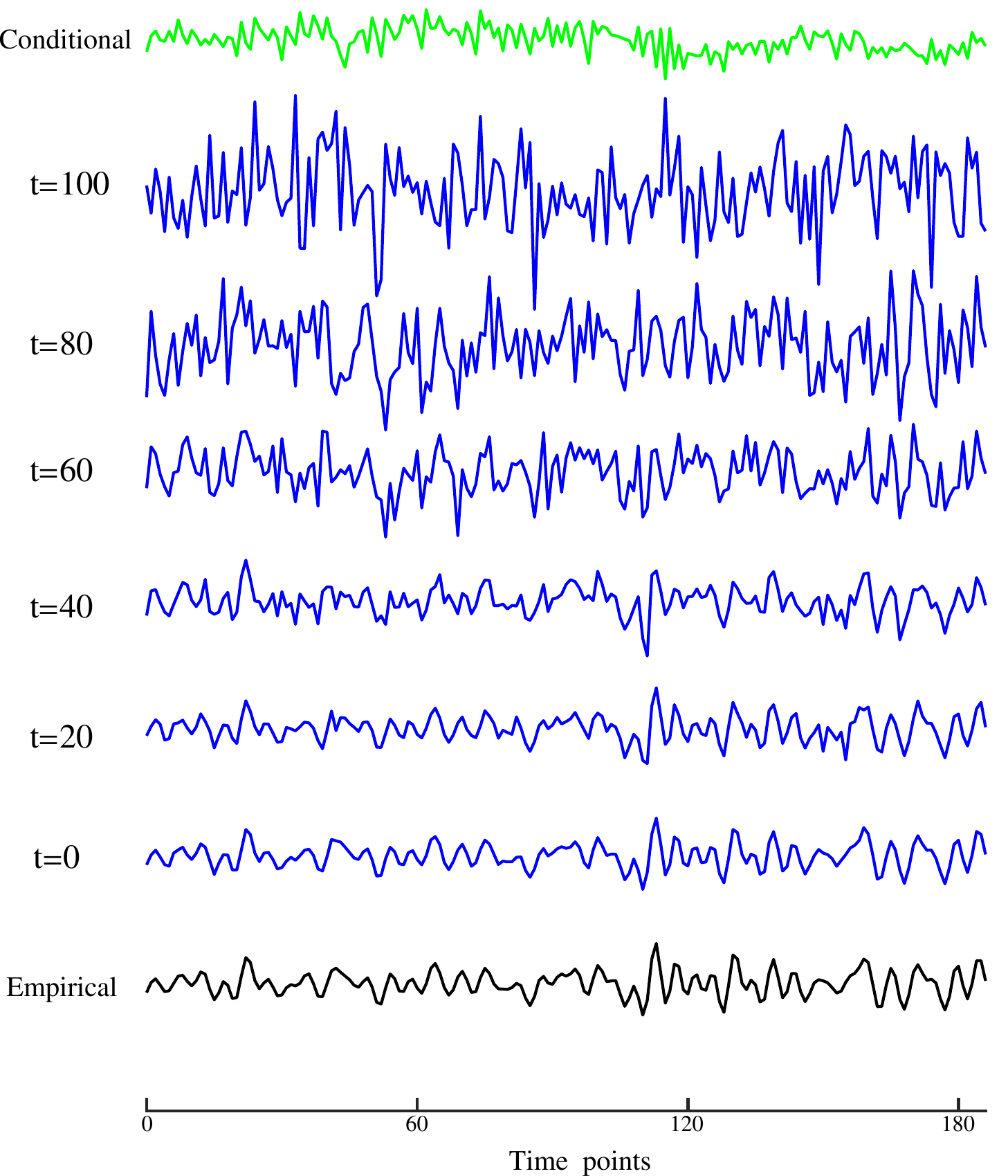}
	\caption{An example of denoising the conditional sample at different steps for the first ROI in the AAL90 atlas. The top arrow is the conditional time series calculated by the non-parametric dot product, and the bottom row is the empirical time series computed by the GRETNA software.\label{fig4}}
\end{figure}

\begin{figure}[htbp]
	\centering
	\includegraphics[width=0.98\columnwidth]{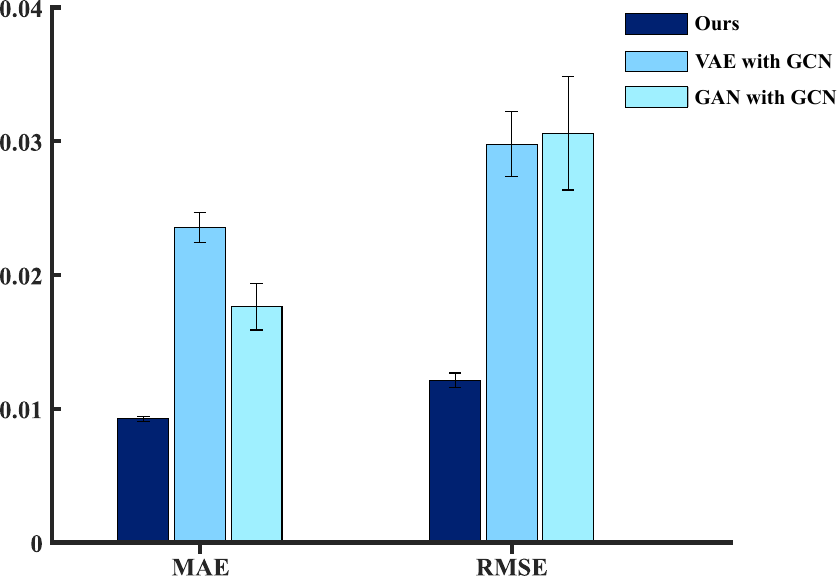}
	\caption{Comparison of reconstructing time series using different methods.\label{fig5}}
\end{figure}

\begin{figure*}[htbp]
	\centering
	\includegraphics[width=0.98\textwidth]{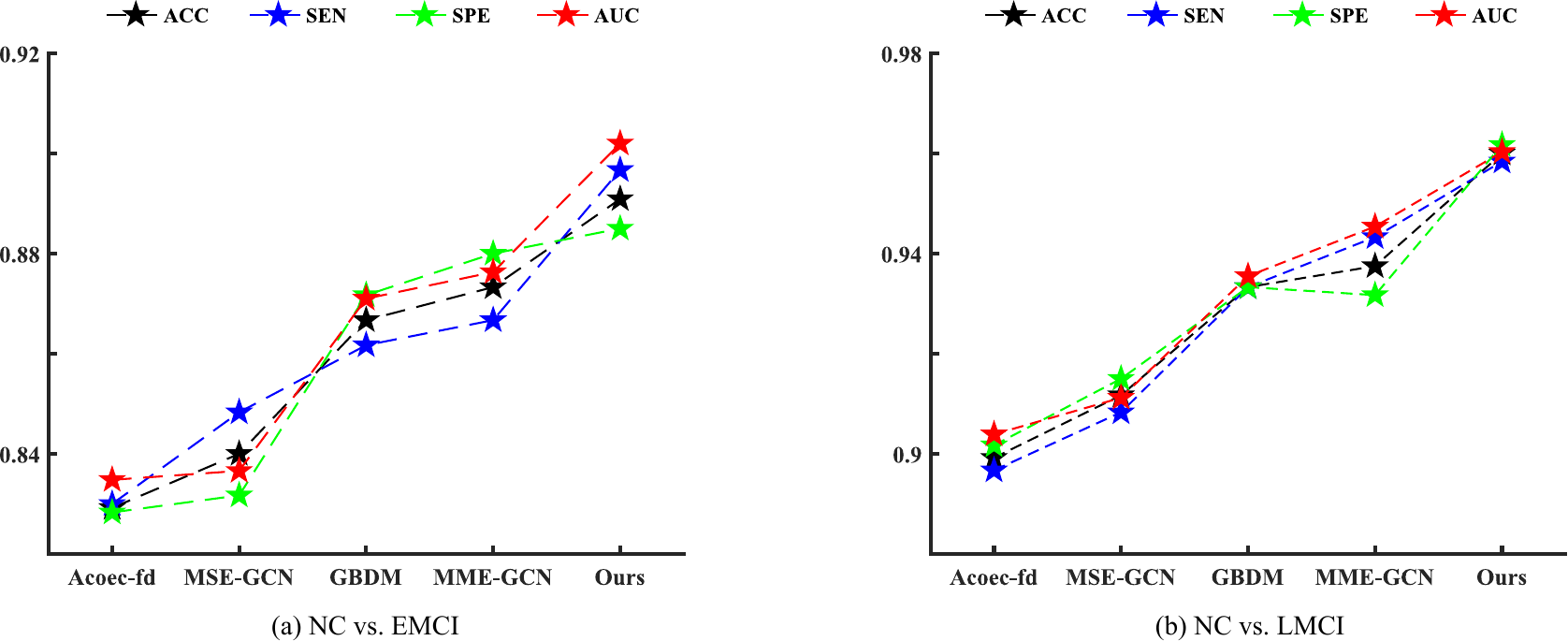}
	\caption{Mean prediction performance of binary classification tasks on the ADNI dataset. Four competing models are compared with our model using fMRI and DTI. \label{fig6}}
\end{figure*}

\begin{table*}[hbp]
\centering
\renewcommand\arraystretch{1.4}
\setlength{\abovecaptionskip}{0pt}%
\setlength{\belowcaptionskip}{10pt}%
\caption{Classification performance comparison of generated brain connectivity using different methods.(mean$\pm$std\%)\label{tab0}}
\begin{tabular}{c|cc|cccc|cccc}
\hline
\multirow{2}{*}{Connectivity type} & \multirow{2}{*}{Method} & \multirow{2}{*}{Modality} & \multicolumn{4}{c|}{NC vs. EMCI}                                                                                                                                                                                                              & \multicolumn{4}{c}{NC vs. LMCI}                                                                                                                                                                                                               \\
                                   &                         &                           & ACC                                                       & SEN                                                       & SPE                                                       & AUC                                                       & ACC                                                       & SEN                                                       & SPE                                                       & AUC                                                       \\ \hline
                                   & MSE-GCN\cite{yu2020multi}                 & fMRI+DTI                  & \begin{tabular}[c]{@{}c@{}}84.00\\ $\pm$0.50\end{tabular} & \begin{tabular}[c]{@{}c@{}}84.83\\ $\pm$1.39\end{tabular} & \begin{tabular}[c]{@{}c@{}}83.17\\ $\pm$1.74\end{tabular} & \begin{tabular}[c]{@{}c@{}}83.66\\ $\pm$2.13\end{tabular} & \begin{tabular}[c]{@{}c@{}}91.17\\ $\pm$1.00\end{tabular} & \begin{tabular}[c]{@{}c@{}}90.83\\ $\pm$1.54\end{tabular} & \begin{tabular}[c]{@{}c@{}}91.50\\ $\pm$2.93\end{tabular} & \begin{tabular}[c]{@{}c@{}}91.12\\ $\pm$1.19\end{tabular} \\ \cline{2-11}
Undirected                         & GBDM\cite{zhang2021deep}                    & fMRI+DTI                  & \begin{tabular}[c]{@{}c@{}}86.67\\ $\pm$0.53\end{tabular} & \begin{tabular}[c]{@{}c@{}}86.17\\ $\pm$1.67\end{tabular} & \begin{tabular}[c]{@{}c@{}}87.17\\ $\pm$1.30\end{tabular} & \begin{tabular}[c]{@{}c@{}}87.10\\ $\pm$1.22\end{tabular} & \begin{tabular}[c]{@{}c@{}}93.33\\ $\pm$0.53\end{tabular} & \begin{tabular}[c]{@{}c@{}}93.33\\ $\pm$1.29\end{tabular} & \begin{tabular}[c]{@{}c@{}}93.33\\ $\pm$1.49\end{tabular} & \begin{tabular}[c]{@{}c@{}}93.55\\ $\pm$1.28\end{tabular} \\ \cline{2-11}
                                   & MME-GCN\cite{liu2022enhanced}                 & fMRI+DTI                  & \begin{tabular}[c]{@{}c@{}}87.33\\ $\pm$0.50\end{tabular} & \begin{tabular}[c]{@{}c@{}}86.67\\ $\pm$1.29\end{tabular} & \begin{tabular}[c]{@{}c@{}}88.00\\ $\pm$1.25\end{tabular} & \begin{tabular}[c]{@{}c@{}}87.63\\ $\pm$1.24\end{tabular} & \begin{tabular}[c]{@{}c@{}}93.75\\ $\pm$0.56\end{tabular} & \begin{tabular}[c]{@{}c@{}}94.33\\ $\pm$1.33\end{tabular} & \begin{tabular}[c]{@{}c@{}}93.17\\ $\pm$1.17\end{tabular} & \begin{tabular}[c]{@{}c@{}}94.54\\ $\pm$1.21\end{tabular} \\ \hline
                                   & Acoec-fd\cite{ji2019acoec}                & fMRI+DTI                  & \begin{tabular}[c]{@{}c@{}}82.92\\ $\pm$0.67\end{tabular} & \begin{tabular}[c]{@{}c@{}}83.00\\ $\pm$2.08\end{tabular} & \begin{tabular}[c]{@{}c@{}}82.83\\ $\pm$2.24\end{tabular} & \begin{tabular}[c]{@{}c@{}}83.48\\ $\pm$2.07\end{tabular} & \begin{tabular}[c]{@{}c@{}}89.92\\ $\pm$1.37\end{tabular} & \begin{tabular}[c]{@{}c@{}}89.67\\ $\pm$1.94\end{tabular} & \begin{tabular}[c]{@{}c@{}}90.17\\ $\pm$1.89\end{tabular} & \begin{tabular}[c]{@{}c@{}}90.39\\ $\pm$1.78\end{tabular} \\ \cline{2-11}
Directed                           & EC-RGAN\cite{ji2021estimating}                & fMRI                      & \begin{tabular}[c]{@{}c@{}}85.50\\ $\pm$0.76\end{tabular} & \begin{tabular}[c]{@{}c@{}}86.00\\ $\pm$1.53\end{tabular} & \begin{tabular}[c]{@{}c@{}}85.00\\ $\pm$1.67\end{tabular} & \begin{tabular}[c]{@{}c@{}}86.24\\ $\pm$1.96\end{tabular} & \begin{tabular}[c]{@{}c@{}}92.35\\ $\pm$0.69\end{tabular} & \begin{tabular}[c]{@{}c@{}}92.42\\ $\pm$1.09\end{tabular} & \begin{tabular}[c]{@{}c@{}}92.27\\ $\pm$1.28\end{tabular} & \begin{tabular}[c]{@{}c@{}}92.45\\ $\pm$1.78\end{tabular} \\ \cline{2-11}
                                   & STGCM\cite{zou2022exploring}                   & fMRI                      & \begin{tabular}[c]{@{}c@{}}87.50\\ $\pm$0.64\end{tabular} & \begin{tabular}[c]{@{}c@{}}86.50\\ $\pm$1.38\end{tabular} & \begin{tabular}[c]{@{}c@{}}88.50\\ $\pm$1.17\end{tabular} & \begin{tabular}[c]{@{}c@{}}87.81\\ $\pm$1.68\end{tabular} & \begin{tabular}[c]{@{}c@{}}94.33\\ $\pm$0.73\end{tabular} & \begin{tabular}[c]{@{}c@{}}93.67\\ $\pm$1.63\end{tabular} & \begin{tabular}[c]{@{}c@{}}95.00\\ $\pm$1.29\end{tabular} & \begin{tabular}[c]{@{}c@{}}94.11\\ $\pm$1.40\end{tabular} \\ \cline{2-11}
                                   & Ours                    & fMRI+DTI                  & {\bfseries\begin{tabular}[c]{@{}c@{}}89.09\\ $\pm$0.45\end{tabular}} & {\bfseries \begin{tabular}[c]{@{}c@{}}89.67\\ $\pm$1.00\end{tabular}} & \begin{tabular}[c]{@{}c@{}}88.50\\ $\pm$0.50\end{tabular} & {\bfseries\begin{tabular}[c]{@{}c@{}}90.20\\ $\pm$1.05\end{tabular}} & {\bfseries \begin{tabular}[c]{@{}c@{}}96.00\\ $\pm$0.50\end{tabular}} & {\bfseries \begin{tabular}[c]{@{}c@{}}95.84\\ $\pm$1.12\end{tabular}} & {\bfseries \begin{tabular}[c]{@{}c@{}}96.17\\ $\pm$0.77\end{tabular}} & {\bfseries \begin{tabular}[c]{@{}c@{}}96.03\\ $\pm$0.50\end{tabular}} \\ \hline
\end{tabular}
\end{table*}

\subsection{Training Settings and Evaluation Metrics}
In our experiment, the proposed model aims to transform the four-dimensional fMRI into ROI-based time series and an effective connectivity network. We set the total diffusing steps as $100$, and the sampling parameters are set as: $T_n=10, m=10$. The $L_0$ is 2. The sparsity constraint parameter $\omega$ is $2.5$. The code is written with Python tools and runs on the Ubuntu 18.04 system. The total number of epochs is 800. The Adam optimizer is adopted to update the model's weights with a learning rate of 0.001.

The results are evaluated using a 5-fold cross-validation strategy. For the reconstruction task, we utilize the mean absolute error (MAE) and the root mean square error (RMSE) to estimate the model's performance. For the classification tasks (i.e., NC vs. EMCI, NC vs. LMCI), the four commonly used metrics are considered: the prediction accuracy (ACC), the patient sensitivity (SEN), the healthy specificity (SPE), and the area under the receiver operating characteristic curve (AUC).

\subsection{Denoising and Prediction Results}
The output of the proposed model is the denoised time series and the effective connectivity. We evaluated the performance of the generated time series. The details of generating one ROI-based time series by our model are displayed in Fig.~\ref{fig4}. As the time steps decrease, the Gaussian noise is gradually transformed into a clean time series. We select two other generative frameworks for comparison, including the variational autoencoder (VAE) and the generative adversarial network (GAN). These two frameworks accept both SC and rough time series by designing graph convolutional networks (GCN). As shown in Fig.~\ref{fig5}, the VAE-based or GAN-based models have larger reconstructed errors, while our model achieves the minimum mean value for MAE and RMSE with 0.0092 and 0.0121, respectively. The standard error of our model is also lower than that of the other two models.

\begin{figure}[h]
	\centering
	\includegraphics[width=0.98\columnwidth]{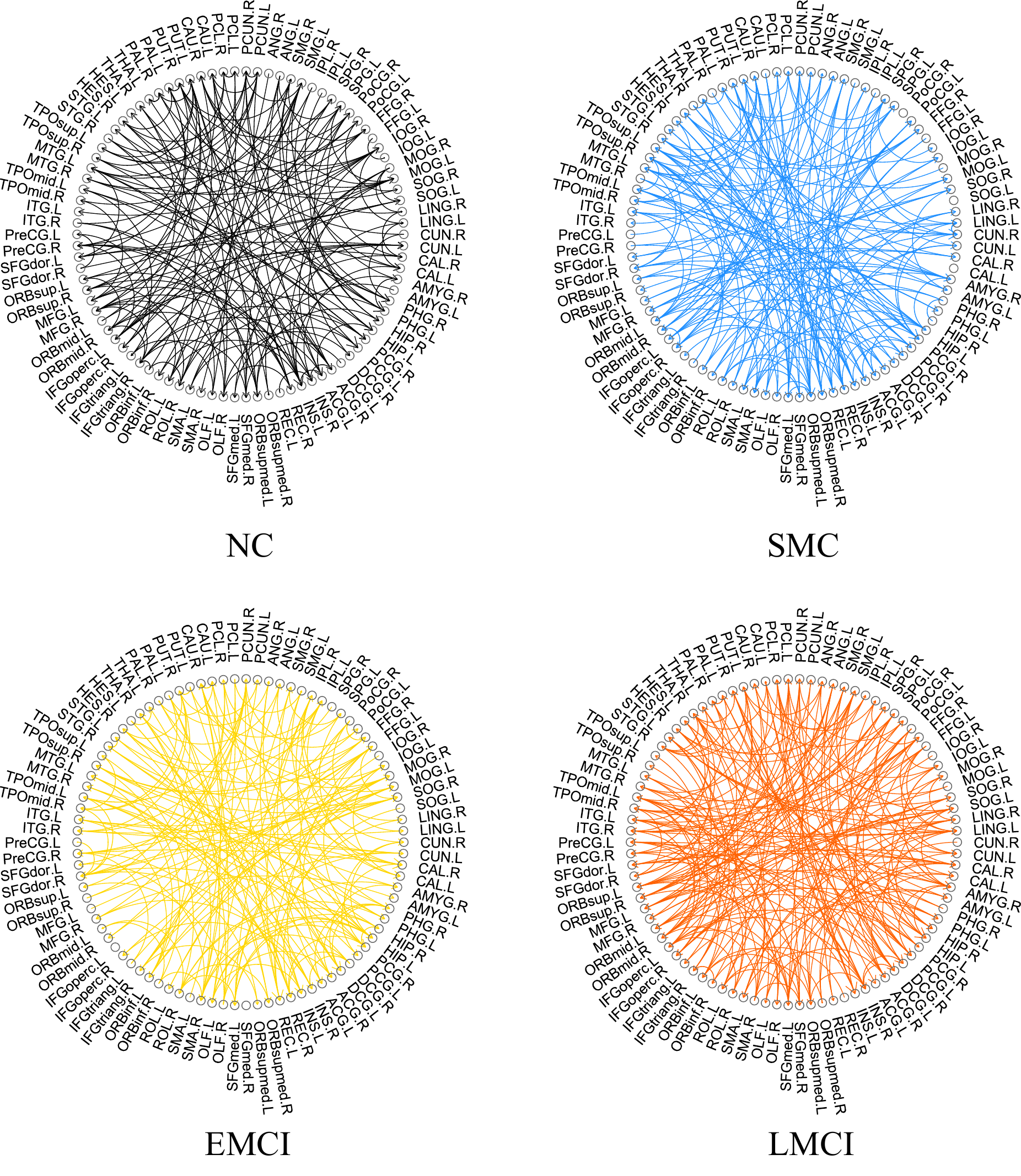}
	\caption{Examples of the effective connectivities generated at four different stages. The 90 brain regions are arranged counterclockwise.\label{fig7}}
\end{figure}

To estimate the quality of generated ECs, we conducted two binary classification tasks (i.e., NC vs. EMCI, and NC vs. LMCI). On the one hand, we computed the classification performance of undirected connectivities generated by different methods. These three methods are (1) the MSE-GCN method \cite{yu2020multi}, (2) the GBDM method \cite{zhang2021deep}, and (3) the MME-GCN method \cite{liu2022enhanced}. They fused fMRI and DTI to generate undirected connectivities, which are sent to the BrainNetCNN classifier \cite{kawahara2017brainnetcnn} for prediction.
On the other hand, we evaluated the classification performance of directed connectivities using four methods. The first one (Acoec-fd\cite{ji2019acoec}) is a shallow learning model that fused the fMRI and DTI to generate effective connectivities (directed connectivities). The second one (EC-RGAN\cite{ji2021estimating}) is a generative model that combined recurrent neural networks and GAN to estimate effective connectivity from fMRI. The third one (STGCM\cite{zou2022exploring}) estimated effective connectivities from fMRI by designing spatiotemporal graph convolutional models. This is different from the previous two models since it is a deep learning-based and non-generative model. The fourth one is our diffusion-based model using fMRI and DTI.
To compare the stability of different models, we first randomly split the datasets using a 5-fold cross-validation strategy and computed the four metrics (ACC, SEN, SPE, and AUC) for each model. Then we repeated the above computation 10 times. The mean and standard deviation values of all the compared models are shown in Tab.~\ref{tab0}. It shows that the classification results of directed connectivities using a single modality achieve comparable performance than those of undirected connectivities using bi-modality. As shown in Fig.~\ref{fig6}, when the model is input with both fMRI and DTI, our model achieves the best performance. This indicates that directed connectivities generated by ours carried more MCI-related information than undirected connectivities generated by other models.
Among the seven models, our model achieves the best classification performance with a mean ACC value of 89.09\%, a mean SEN value of 89.67\%, and a mean AUC value of 90.20\% for NC vs. EMCI.
For NC vs. LMCI, the best mean values of ACC, SEN, SPE, and AUC are 96.00\%, 95.84\%, 96.17\%, and 96.03\%, respectively.
Fig.~\ref{fig7} shows the generated ECs of four subjects by the proposed model. Each EC represents one of the stages from NC to LMCI.

\begin{figure}[htbp]
	\centering
	\includegraphics[width=0.98\columnwidth]{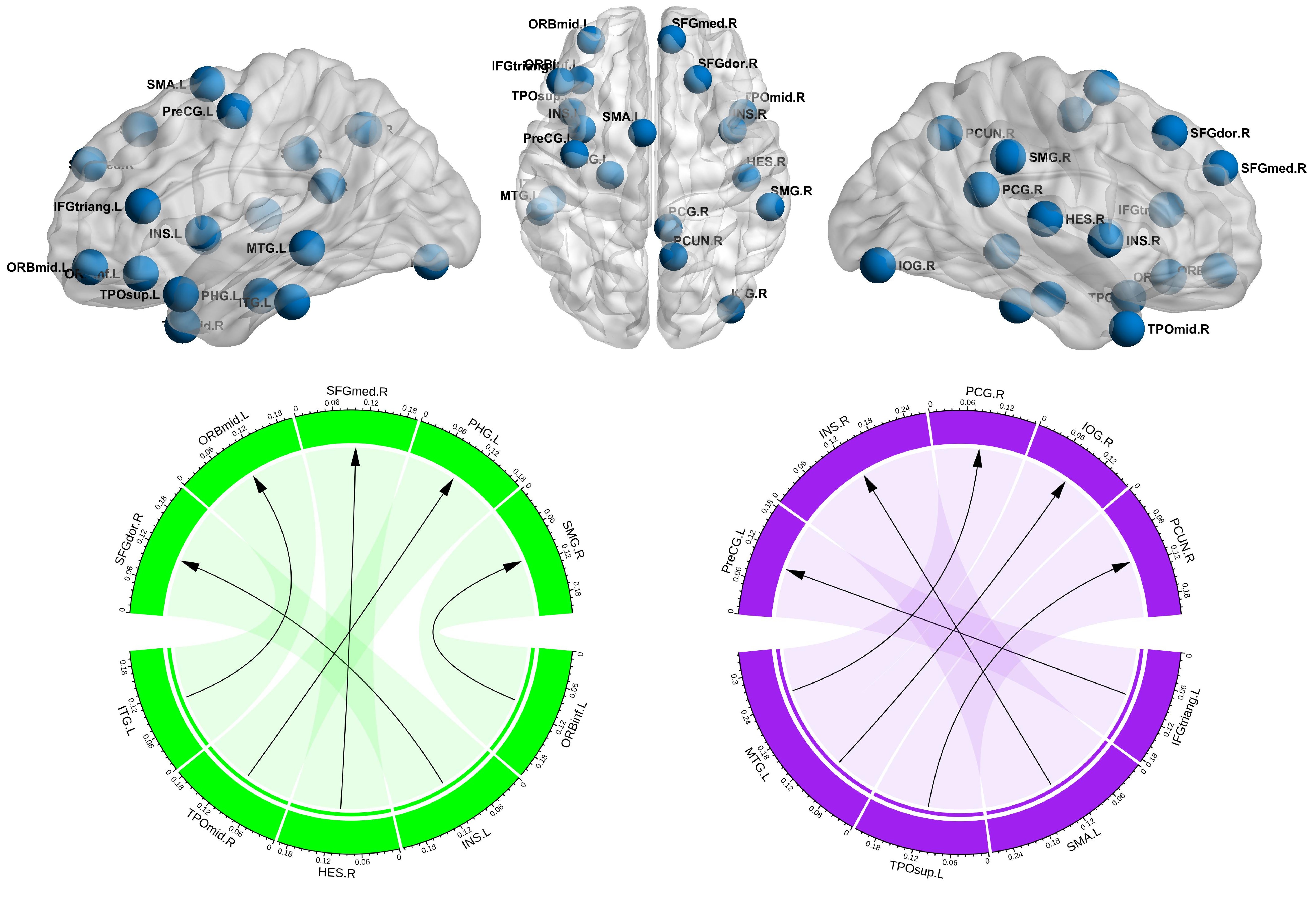}
	\caption{Visualization of the top five altered connections from NC to SMC. The green color means enhanced connections, and the purple color means diminished connections. \label{fig8}}
\end{figure}
\begin{figure}[htbp]
	\centering
	\includegraphics[width=0.98\columnwidth]{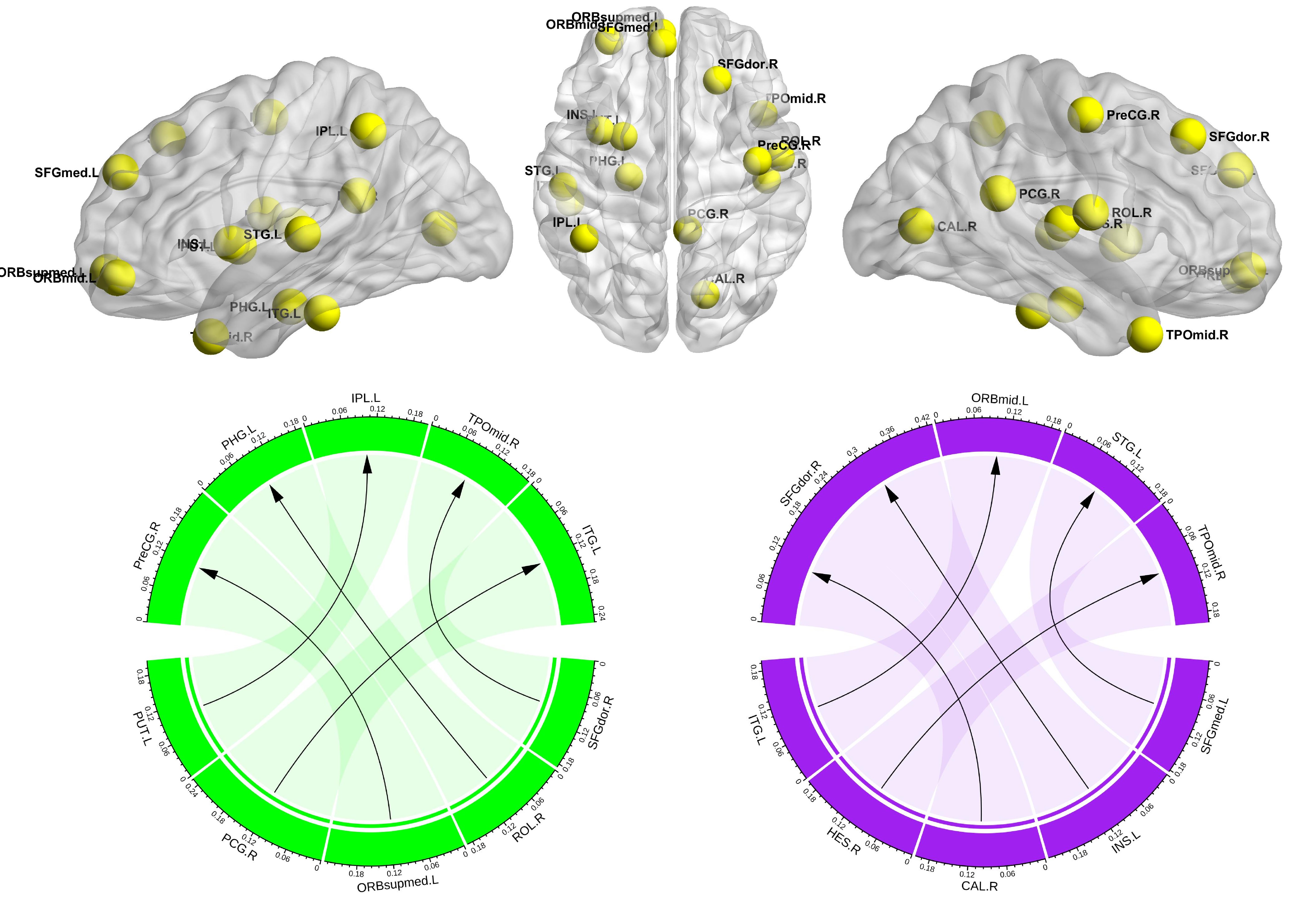}
	\caption{Visualization of the top five enhanced and diminished connections from SMC to EMCI. \label{fig9}}
\end{figure}

\begin{figure}[htbp]
	\centering
	\includegraphics[width=0.98\columnwidth]{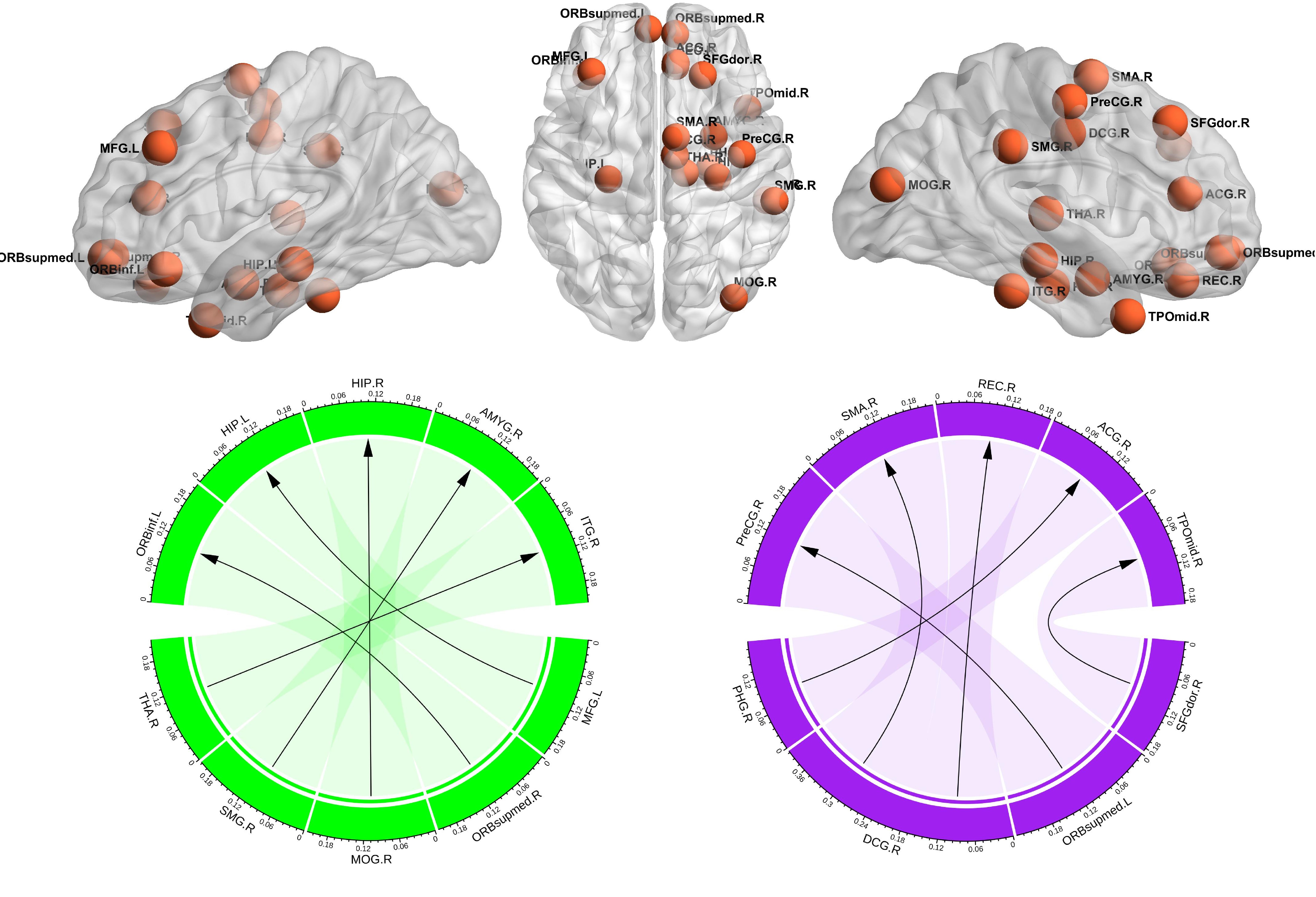}
	\caption{Visualization of the top five enhanced and diminished connections from EMCI to LMCI. \label{fig10}}
\end{figure}

\subsection{Analysis of Effective Connectivity}
During disease progression, the effective connectivity can show abnormal patterns. We call these patterns altered connections. As mentioned in the previous studies \cite{zuo2023brain}, we compute the mean EC for each group (i.e., NC, SMC, EMCI, and LMCI).
The altered connections can be calculated by subtracting the averaged EC at the former stage (NC/SMC/EMCI) from the averaged EC at the latter stage (SMC/EMCI/LMCI).
The positive value in altered connections means enhanced connection, and the negative value means diminished connection. To obtain the top 5 altered connections, we sort the positive and negative values, respectively.
As shown in Fig.~\ref{fig8}, from NC to SMC, the top part displays the distribution of these altered connection-related brain regions, and the bottom part is the altered connections. Fig.~\ref{fig9} and Fig.~\ref{fig10} show the top ten altered connections for SMC-to-EMCI and EMCI-to-LMCI, respectively.

\begin{table}[htbp]
	\centering
	\renewcommand\arraystretch{1.4}
	\setlength{\abovecaptionskip}{0pt}%
	\setlength{\belowcaptionskip}{10pt}%
	\caption{Top five bidirectionally altered connections for three scenarios.}\label{table1}
	\resizebox{\linewidth}{!}{
		\begin{tabular}{ccc}
			\hline
			\multicolumn{1}{l}{}           & ROIs index & ROI names             \\ \hline
			\multirow{5}{*}{NC vs. SMC}    & 19, 30     & SMA.L $\leftrightarrow$ INS.R         \\
			& 15, 64     & ORBinf.L $\leftrightarrow$ SMG.R      \\
			& 1, 13      & PreCG.L $\leftrightarrow$ IFGtriang.L \\
			& 21, 69     & OLF.L $\leftrightarrow$ PCL.L         \\
			& 26, 60     & ORBsupmed.R $\leftrightarrow$ SPG.R   \\ \hline
			\multirow{5}{*}{SMC vs. EMCI}  & 36, 89     & PCG.R $\leftrightarrow$ ITG.L         \\
			& 4, 29      & SFGdor.R $\leftrightarrow$ INS.L      \\
			& 2, 25      & PreCG.R $\leftrightarrow$ ORBsupmed.L \\
			& 5, 20      & ORBsup.L $\leftrightarrow$ SMA.R      \\
			& 27, 31     & REC.L $\leftrightarrow$ ACG.L         \\ \hline
			\multirow{5}{*}{EMCI vs. LMCI} & 4, 88      & SFGdor.R $\leftrightarrow$ TPOmid.R   \\
			& 37, 61     & HIP.L $\leftrightarrow$ IPL.L         \\
			& 20, 78     & SMA.R $\leftrightarrow$ THA.R         \\
			& 2, 25      & PreCG.R$\leftrightarrow$ ORBsupmed.L \\
			& 68, 87     & PCUN.R $\leftrightarrow$ TPOmid.L     \\ \hline
	\end{tabular}}
\end{table}

\begin{figure*}[t!]
	\centering
	\includegraphics[width=\textwidth]{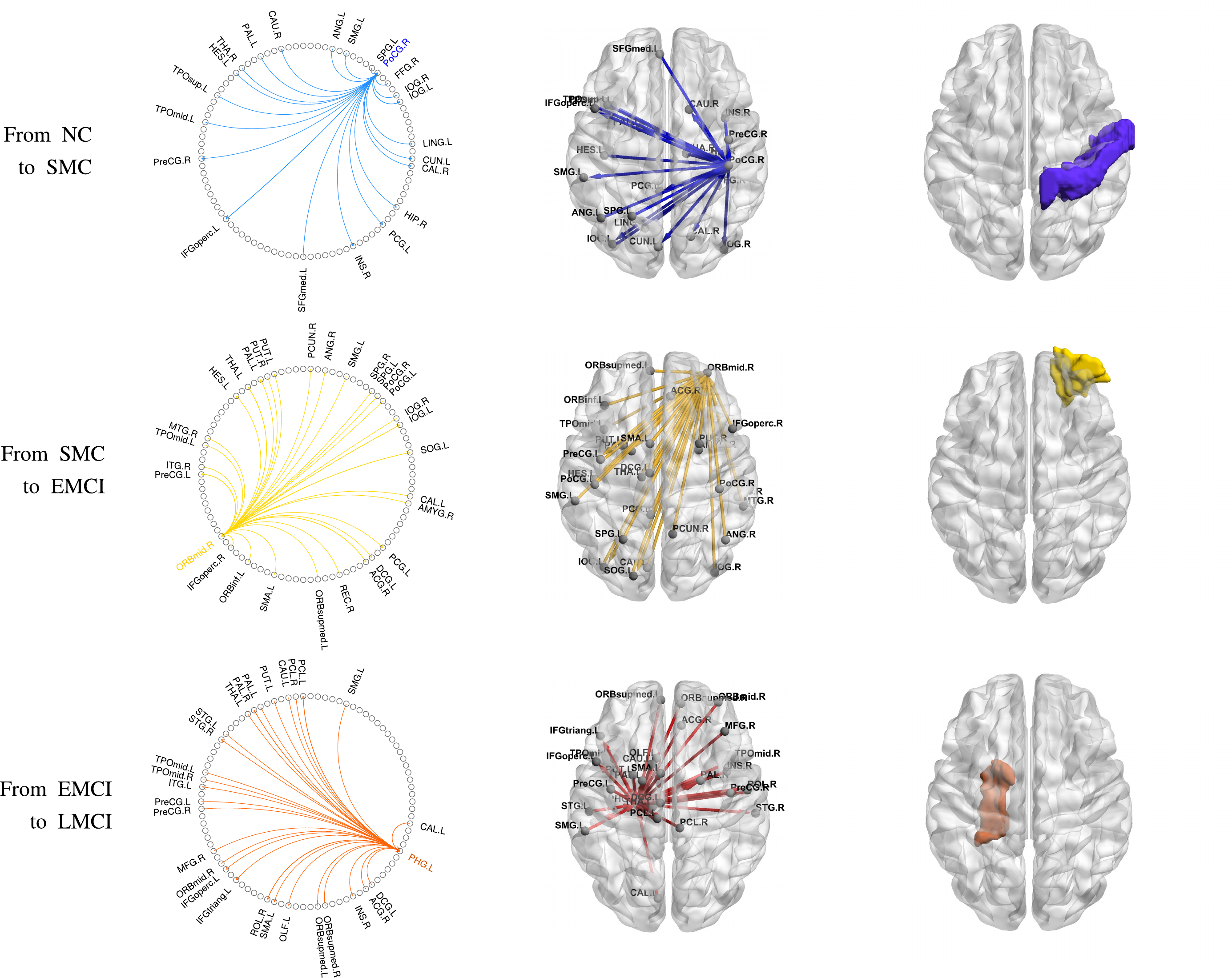}
	\caption{Visualization of the top ROI-related altered connections. The top ROI is calculated by first summing the absolute altered connection strength for each ROI and then sorting all the ROIs' strength in descending order. The three brain regions are PoCG.R, ORBmid.R, and PHG.L for NC vs. SMC, SMC vs. EMCI, and EMCI vs. LMCI, respectively. \label{fig11}}
\end{figure*}

\begin{figure*}[t]
	\centering
	\includegraphics[width=\textwidth]{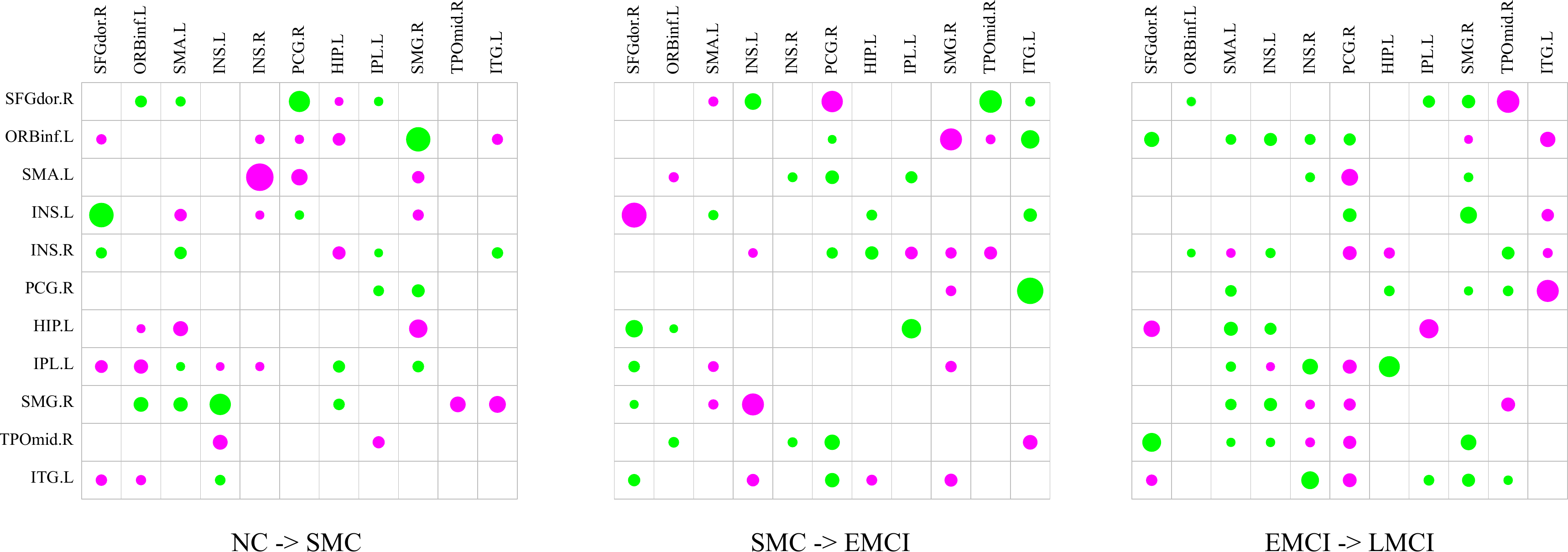}
	\caption{Visualization of the selected 11 ROI-related bidirectionally altered connections. The magenta and green colors represent the diminished and enhanced connections, respectively. The circle size indicates the relative connection strength. \label{fig12}}
\end{figure*}

\begin{figure*}[htbp]
	\centering
	\includegraphics[width=\textwidth]{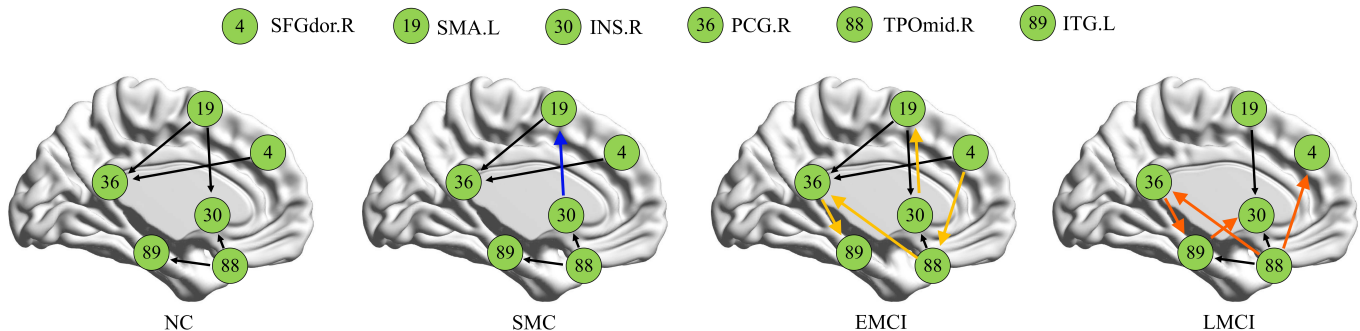}
	\caption{Display of effective connectivity among six representative ROIs by projecting them onto the sagittal plane of the brain at four different stages The black arrow means the effective connectivity at NC, and the colored arrows means altered connections. \label{fig13}}
\end{figure*}

To analyze the important brain regions that contribute most to MCI, we ignore the sign of each altered connection and sum the altered connection strength for each brain region. After sorting the summed connection strength in descending order, the top five ROIs are the potential biomarkers in the MCI progression. Fig.~\ref{fig11} shows the top brain region and its related altered connections in three scenarios. The top five ROIs for NC vs. SMC are PoCG.R, AMYG.L, ROL.R, LING.R and FFG.L. From SMC to EMCI, the five important ROIs are ORBmid.R, SMG.R, IFGoperc.L, SOG.R and ORBsup.R. The five important ROIs between the EMCI and LMCI groups are PHG.L, SFGdor.L, SMG.L, SOG.L and IFGtriang.L. These identified ROIs are partly consistent with previous studies \cite{lei2020self, song2021graph}. Furthermore, we analyze the bidirectionally altered connections among different groups.
For each bidirectionally altered connection, we ignore the sign and sum the values of each bidirectionally altered connection's strength. The subsequent summed values are sorted in descending order.
The calculated top five bidirectional altered connections are shown in Tab.~\ref{table1}. To compare these connection characteristics at different stages, we select 4 ROIs that are in the top two bidirectional altered connections for each scenario and present all these ROIs in Fig.~\ref{fig12}. The connection strength among these 11 ROIs is normalized into the range $-1 \sim 1$ for comparative analysis. The absolute values below 0.1 are not displayed. The number of enhanced connections (green color) tends to increase compared with that of the diminished connections (magenta color) as the disease progresses. This characteristic can be explained by the compensatory mechanism at the MCI stage \cite{qi2010impairment}. In addition, we select the top bidirectional altered connection for each scenario and obtain six brain regions for three scenarios. Fig.~\ref{fig13} shows the effective connectivity characteristic changing among six brain regions during the disease progression. For visualization convenience, all brain regions are projected onto the sagittal plane. Compared with NC, more enhanced connections are founded at MCI-related stages than at SMC. The data of the experimental results is available on the website (https://github.com/zookhbue/BDHT).

\subsection{Ablation Study}
The proposed BDHT model can denoise the fMRI into effective connectivity. To prove the effectiveness of the designed network structure, we studied two variants of BDHT to evaluate the prediction performance of NC vs. EMCI. One variant (BDHT w/o U-shape) removes the U-shaped structure in the BDHT, which means the DD and DH are included in our model; the other variant (BDHT w/o Transformer) removes the SMA and TMA in the GraphConFormer block. The prediction results are listed in Tab.~\ref{tab2}. The BDHT without the transformer shows the worst classification performance, which drops at least 4 percentage points on the ACC value. While removing the U-shape structure in the BDHT drops about 2 percent on the ACC calculation, it probably means that the spatiotemporal features carry much more MCI-related information than the multi-scale features.

\begin{table}[htbp]
	\centering
	\renewcommand\arraystretch{1.4}
	\setlength{\abovecaptionskip}{0pt}%
	\setlength{\belowcaptionskip}{10pt}%
	\caption{Impact of the BDHT structure on the prediction performance.(\%)\label{tab2}}
	\resizebox{\linewidth}{!}{
		\begin{tabular}{ccccc}
			\hline
			 & ACC   & SEN   & SPE   & AUC   \\ \hline
			BDHT w/o U-shape                & 85.00 & 86.67 & 83.33 & 85.14 \\
			BDHT w/o Tranformer             & 87.50 & 86.67 & 88.33 & 87.67 \\
			proposed BDHT                   & 89.17 & 90.00 & 88.33 & 91.33 \\ \hline
	\end{tabular}}
\end{table}

\section{Discussion}
\label{s5}
Our work is a unifying generative framework to estimate brain effective connectivity for MCI analysis. This framework provides a new insight to transform four-dimensional fMRI into effective connectivity using the generative artificial intelligence. The proposed model can not only diagnose brain neurological diseases (including MCI), but also identify altered directional connections to understand pathogenesis of cognitive disease for developing targeted therapeutic interventions. Previous studies obtained the ROI-based time series from fMRI by using certain software. A few of them focus on combining the ROI-based time series using a deep learning model. Our model is methodologically feasible. The ROI mask of the AAL90 atlas is prior knowledge that can partition a three-dimensional volume into 90 parts for each time point. This step has no learnable parameters. The mean pixel value of each ROI can generate rougth ROI-based time series. This rough sample has unknown noise compared with the empirical sample (using the software toolbox), which can be treated as a condition in the reverse direction of DDPM. Furthermore, the structural connectivity can be used to extract topological characteristics in the reverse process. Meanwhile, the causal relationships between brain regions can be explored using the SEM method. Fig.~\ref{fig5} shows that the proposed BDHT has the potential to replace the software toolbox in preprocessing functional MRI.

The experimental results show important unidirectional and bidirectionally altered connections. The frequently appearing brain regions are the superior frontal gyrus dorsolateral, inferior frontal gyrus triangular part, superior frontal gyrus medial orbital, parahippocampal gyrus, supramarginal gyrus, and temporal pole middle temporal gyrus. These brain regions are reported to be correlated with MCI \cite{wang2021deficit,chung2022association,lin2019multiparametric,bozzali2011anatomical}. For example, the parahippocampal gyrus's main functions include creating memories and recalling visual scenes. Damage to this gyrus in patients disrupts the memory-related neural circuit and causes cognitive decline. The supramarginal gyrus controls language expression in the neural center; abnormal connections with it cause language dysfunction and severely affect social activities.
Fig.~\ref{fig13} gives the EC changing characteristics among six representative ROIs from NC to LMCI. The SMC stage shows one enhanced and one diminished connection compared with the NC stage. As the progression comes at EMCI, four enhanced and one diminished connections are emerging. When LMCI occurs, four enhanced and two diminished connections occur. This phenomenon can be explained by the following: SMC shows slight changes in the brain network; EMCI compensates the whole brain network for daily activities; and LMCI causes further damage to the brain functions. Therefore, much more attention should be paid to the EMCI stage.

The main limitation of this work is that it ignores the dynamic changes in characteristics during cognitive progression. As the functional time series carries much more underlying information about the MCI, constructing dynamic effective connectivities can uncover the pathological mechanisms of the MCI and make the results more interpretable. In the future, we will bridge the dynamic causal relationships among brain regions for clinical applications.

\section{Conclusion}
\label{s6}
This paper proposes a brain diffuser with hierarchical transformer (BDHT) to construct effective connectivity from four-dimensional fMRI for MCI analysis. The BDHT leverages structural connectivity to guide the denoising process using the generative artificial intelligence. The designed U-shape structure learns multi-scale features in topological space for denoising enhancement. Meanwhile, the GraphConFormer block, consisting of the multi-head attention and graph convolutional network, improves the structure-function fusion effect and boosts noise removal ability. The proposed model was evaluated on the ADNI dataset and achieves superior denoising quality and prediction performance than related models. The BDHT can identify important brain regions and altered connections for MCI diagnosis. The identified altered connections may have the potential to be biomarkers for MCI treatment.

\section*{Data Availability Statement}
The experimental dataset in this article is readily available in the [Alzheimer's Disease Neuroimaging Initiative](\href{https://adni.loni.usc.edu}{https://adni.loni.usc.edu}).

\section*{Acknowledgments}
This research is supported partly by the Natural Science Foundation of Hubei Province (No. 2023AFB004), partly by the National Natural Science Foundation of China (No. 61872351).


%
%
%
%
%
%
%
%
%


\begin{thebibliography}{1}
\bibliographystyle{IEEEtran}

\bibitem{gupta2016traumatic}
R.~Gupta and N.~Sen, ``Traumatic brain injury: a risk factor for
  neurodegenerative diseases,'' \emph{Reviews in the Neurosciences}, vol.~27,
  no.~1, pp. 93--100, 2016.

\bibitem{lei2022predicting}
B.~Lei, E.~Liang, M.~Yang, P.~Yang, F.~Zhou, E.-L. Tan, Y.~Lei, C.-M. Liu,
  T.~Wang, X.~Xiao \emph{et~al.}, ``Predicting clinical scores for
  alzheimer's disease based on joint and deep learning,'' \emph{Expert
  Systems with Applications}, vol. 187, p. 115966, 2022.

\bibitem{myszczynska2020applications}
M.~A. Myszczynska, P.~N. Ojamies, A.~M. Lacoste, D.~Neil, A.~Saffari, R.~Mead,
  G.~M. Hautbergue, J.~D. Holbrook, and L.~Ferraiuolo, ``Applications of
  machine learning to diagnosis and treatment of neurodegenerative diseases,''
  \emph{Nature Reviews Neurology}, vol.~16, no.~8, pp. 440--456, 2020.

\bibitem{deshpande2011instantaneous}
G.~Deshpande, P.~Santhanam, and X.~Hu, ``Instantaneous and causal connectivity
  in resting state brain networks derived from functional mri data,''
  \emph{Neuroimage}, vol.~54, no.~2, pp. 1043--1052, 2011.

\bibitem{park2018dynamic}
H.-J. Park, K.~J. Friston, C.~Pae, B.~Park, and A.~Razi, ``Dynamic effective
  connectivity in resting state fmri,'' \emph{NeuroImage}, vol. 180, pp.
  594--608, 2018.

\bibitem{zhong2014altered}
Y.~Zhong, L.~Huang, S.~Cai, Y.~Zhang, K.~M. von Deneen, A.~Ren, J.~Ren,
  A.~D.~N. Initiative \emph{et~al.}, ``Altered effective connectivity patterns
  of the default mode network in alzheimer's disease: an fmri study,''
  \emph{Neuroscience letters}, vol. 578, pp. 171--175, 2014.


\bibitem{xia2023structure}
Z.~Xia, T.~Zhou, S.~Mamoon, A.~Alfakih, and J.~Lu, ``A structure-guided
  effective and temporal-lag connectivity network for revealing brain disorder
  mechanisms,'' \emph{IEEE Journal of Biomedical and Health Informatics}, 2023.



\bibitem{lei2021diagnosis}
B.~Lei, S.~Yu, X.~Zhao, A.~F. Frangi, E.-L. Tan, A.~Elazab, T.~Wang, and
  S.~Wang, ``Diagnosis of early alzheimer's disease based on dynamic high
  order networks,'' \emph{Brain imaging and behavior}, vol.~15, pp. 276--287,
  2021.

\bibitem{gong2023generative}
C.~Gong, C.~Jing, X.~Chen, C.~M. Pun, G.~Huang, A.~Saha, M.~Nieuwoudt, H.-X.
  Li, Y.~Hu, and S.~Wang, ``Generative ai for brain image computing and brain
  network computing: a review,'' \emph{Frontiers in Neuroscience}, vol.~17, p.
  1203104, 2023.

\bibitem{zuo2023alzheimer}
Q.~Zuo, Y.~Shen, N.~Zhong, C.~P. Chen, B.~Lei, and S.~Wang, ``Alzheimer's
  disease prediction via brain structural-functional deep fusing network,''
  \emph{IEEE Transactions on Neural Systems and Rehabilitation Engineering},
  vol.~31, pp. 4601--4612, 2023.

\bibitem{wang2024novel}
L.~Wang, H.~Wang, Y.~Su, F.~Lure, and J.~Li, ``A novel hybrid ordinal learning
  model with health care application,'' \emph{IEEE Transactions on Automation
  Science and Engineering}, 2024.



\bibitem{cao2022brain}
J.~Cao, Y.~Zhao, X.~Shan, H.-l. Wei, Y.~Guo, L.~Chen, J.~A. Erkoyuncu, and
  P.~G. Sarrigiannis, ``Brain functional and effective connectivity based on
  electroencephalography recordings: A review,'' \emph{Human brain mapping},
  vol.~43, no.~2, pp. 860--879, 2022.



\bibitem{cosio2022diagnosis}
R.~Cos{\'\i}o-Guirado, C.~Soriano-Mas, I.~Del~Cerro, M.~Urretavizcaya, J.~M.
  Mench{\'o}n, V.~Soria, C.~Ca{\~n}ete-Mass{\'e}, M.~Per{\'o}-Cebollero, and
  J.~Guardia-Olmos, ``Diagnosis of late-life depression using structural
  equation modeling and dynamic effective connectivity during resting fmri,''
  \emph{Journal of Affective Disorders}, vol. 318, pp. 246--254, 2022.


\bibitem{li2020large}
G.~Li, Y.~Liu, Y.~Zheng, D.~Li, X.~Liang, Y.~Chen, Y.~Cui, P.-T. Yap, S.~Qiu,
  H.~Zhang \emph{et~al.}, ``Large-scale dynamic causal modeling of major
  depressive disorder based on resting-state functional magnetic resonance
  imaging,'' \emph{Human brain mapping}, vol.~41, no.~4, pp. 865--881, 2020.


\bibitem{jiang2021analysing}
X.~Jiang and H.~Wang, ``Analysing effective connectivity of the math-gifted
  brain with nonlinear granger causality,'' in \emph{2021 IEEE 4th Advanced
  Information Management, Communicates, Electronic and Automation Control
  Conference (IMCEC)}, vol.~4.\hskip 1em plus 0.5em minus 0.4em\relax IEEE,
  2021, pp. 1932--1936.


\bibitem{pan2021characterization}
J.~Pan, B.~Lei, Y.~Shen, Y.~Liu, Z.~Feng, and S.~Wang, ``Characterization
  multimodal connectivity of brain network by hypergraph gan for alzheimer's
  disease analysis,'' in \emph{Pattern Recognition and Computer Vision: 4th
  Chinese Conference, PRCV 2021, Beijing, China, October 29--November 1, 2021,
  Proceedings, Part III 4}.\hskip 1em plus 0.5em minus 0.4em\relax Springer,
  2021, pp. 467--478.


\bibitem{zuo2023prior}
Q.~Zuo, H.~Wu, C.~P. Chen, B.~Lei, and S.~Wang, ``Prior-guided adversarial
  learning with hypergraph for predicting abnormal connections in alzheimer's
  disease,'' \emph{IEEE Transactions on Cybernetics}, 2024.


\bibitem{scherr2021effective}
M.~Scherr, L.~Utz, M.~Tahmasian, L.~Pasquini, M.~J. Grothe, J.~P. Rauschecker,
  T.~Grimmer, A.~Drzezga, C.~Sorg, and V.~Riedl, ``Effective connectivity in
  the default mode network is distinctively disrupted in alzheimer's
  disease: a simultaneous resting-state fdg-pet/fmri study,'' \emph{Human
  brain mapping}, vol.~42, no.~13, pp. 4134--4143, 2021.



\bibitem{ji2019acoec}
J.~Ji, J.~Liu, A.~Zou, and A.~Zhang, ``Acoec-fd: Ant colony optimization for
  learning brain effective connectivity networks from functional mri and
  diffusion tensor imaging,'' \emph{Frontiers in Neuroscience}, vol.~13, p.
  1290, 2019.


\bibitem{wu2023multi}
H.~Wu, B.~Zhang, J.~Pan, and J.~Qin, ``Multi-level object-aware guidance
  network for biomedical image segmentation,'' \emph{IEEE Transactions on
  Automation Science and Engineering}, 2023.

\bibitem{wang2020ensemble}
S.~Wang, X.~Wang, Y.~Shen, B.~He, X.~Zhao, P.~W.-H. Cheung, J.~P.~Y. Cheung,
  K.~D.-K. Luk, and Y.~Hu, ``An ensemble-based densely-connected deep learning
  system for assessment of skeletal maturity,'' \emph{IEEE Transactions on
  Systems, Man, and Cybernetics: Systems}, vol.~52, no.~1, pp. 426--437, 2020.

\bibitem{hu2020medical}
S.~Hu, W.~Yu, Z.~Chen, and S.~Wang, ``Medical image reconstruction using
  generative adversarial network for alzheimer disease assessment with
  class-imbalance problem,'' in \emph{2020 IEEE 6th international conference on
  computer and communications (ICCC)}.\hskip 1em plus 0.5em minus 0.4em\relax
  IEEE, 2020, pp. 1323--1327.

\bibitem{hu20233}
B.~Hu, C.~Zhan, B.~Tang, B.~Wang, B.~Lei, and S.-Q. Wang, ``3-d brain
  reconstruction by hierarchical shape-perception network from a single
  incomplete image,'' \emph{IEEE Transactions on Neural Networks and Learning
  Systems}, 2023.



\bibitem{yang2023spatial}
R.~Yang, W.~Dai, H.~She, Y.~P. Du, D.~Wu, and H.~Xiong, ``Spatial-temporal dag
  convolutional networks for end-to-end joint effective connectivity learning
  and resting-state fmri classification,'' \emph{arXiv preprint
  arXiv:2312.10317}, 2023.

\bibitem{zhang2023amortization}
Z.~Zhang, Z.~Zhang, J.~Ji, and J.~Liu, ``Amortization transformer for brain
  effective connectivity estimation from fmri data,'' \emph{Brain sciences},
  vol.~13, no.~7, p. 995, 2023.

\bibitem{ji2021estimating}
J.~Ji, J.~Liu, L.~Han, and F.~Wang, ``Estimating effective connectivity by
  recurrent generative adversarial networks,'' \emph{IEEE Transactions on
  Medical Imaging}, vol.~40, no.~12, pp. 3326--3336, 2021.

\bibitem{zou2022exploring}
A.~Zou, J.~Ji, M.~Lei, J.~Liu, and Y.~Song, ``Exploring brain effective
  connectivity networks through spatiotemporal graph convolutional models,''
  \emph{IEEE Transactions on Neural Networks and Learning Systems}, 2022.


\bibitem{ji2023dynamic}
J.~Ji, L.~Han, F.~Wang, and J.~Liu, ``Dynamic effective connectivity learning
  based on non-parametric state estimation and gan,'' \emph{IEEE Transactions
  on Instrumentation and Measurement}, 2023.

\bibitem{liu2024mcan}
J.~Liu, L.~Han, and J.~Ji, ``Mcan: multimodal causal adversarial networks for
  dynamic effective connectivity learning from fmri and eeg data,'' \emph{IEEE
  Transactions on Medical Imaging}, 2024.
\bibitem{10097497}
Z.~Xia, T.~Zhou, S.~Mamoon, A.~Alfakih, and J.~Lu, ``A structure-guided
  effective and temporal-lag connectivity network for revealing brain disorder
  mechanisms,'' \emph{IEEE Journal of Biomedical and Health Informatics},
  vol.~27, no.~6, pp. 2990--3001, 2023.

\bibitem{ji2024metacae}
J.~Ji, Z.~Zhang, L.~Han, and J.~Liu, ``Metacae: Causal autoencoder with
  meta-knowledge transfer for brain effective connectivity estimation,''
  \emph{Computers in Biology and Medicine}, vol. 170, p. 107940, 2024.






\bibitem{zhou2023scgan}
S.~Zhou, U.~J. Islam, N.~Pfeiffer, I.~Banerjee, B.~K. Patel, and A.~S. Iquebal,
  ``Scgan: Sparse countergan for counterfactual explanations in breast cancer
  prediction,'' \emph{IEEE Transactions on Automation Science and Engineering},
  2023.

\bibitem{li2023dls}
W.~Li, C.~Gu, J.~Chen, C.~Ma, X.~Zhang, B.~Chen, and S.~Wan, ``Dls-gan:
  generative adversarial nets for defect location sensitive data
  augmentation,'' \emph{IEEE Transactions on Automation Science and
  Engineering}, 2023.

\bibitem{wang2020diabetic}
S.~Wang, X.~Wang, Y.~Hu, Y.~Shen, Z.~Yang, M.~Gan, and B.~Lei, ``Diabetic
  retinopathy diagnosis using multichannel generative adversarial network with
  semisupervision,'' \emph{IEEE Transactions on Automation Science and
  Engineering}, vol.~18, no.~2, pp. 574--585, 2020.

\bibitem{hu2020brain}
S.~Hu, Y.~Shen, S.~Wang, and B.~Lei, ``Brain mr to pet synthesis via
  bidirectional generative adversarial network,'' in \emph{Medical Image
  Computing and Computer Assisted Intervention--MICCAI 2020: 23rd International
  Conference, Lima, Peru, October 4--8, 2020, Proceedings, Part II 23}.\hskip
  1em plus 0.5em minus 0.4em\relax Springer, 2020, pp. 698--707.

\bibitem{wang2022brain}
S.~Wang, Z.~Chen, S.~You, B.~Wang, Y.~Shen, and B.~Lei, ``Brain stroke lesion
  segmentation using consistent perception generative adversarial network,''
  \emph{Neural Computing and Applications}, vol.~34, no.~11, pp. 8657--8669,
  2022.

\bibitem{ho2020denoising}
J.~Ho, A.~Jain, and P.~Abbeel, ``Denoising diffusion probabilistic models,''
  \emph{Advances in Neural Information Processing Systems}, vol.~33, pp.
  6840--6851, 2020.

\bibitem{nichol2021improved}
A.~Q. Nichol and P.~Dhariwal, ``Improved denoising diffusion probabilistic
  models,'' in \emph{International Conference on Machine Learning}.\hskip 1em
  plus 0.5em minus 0.4em\relax PMLR, 2021, pp. 8162--8171.

\bibitem{park2013structural}
H.-J. Park and K.~Friston, ``Structural and functional brain networks: from
  connections to cognition,'' \emph{Science}, vol. 342, no. 6158, p. 1238411,
  2013.




\bibitem{guo2023shadowdiffusion}
L.~Guo, C.~Wang, W.~Yang, S.~Huang, Y.~Wang, H.~Pfister, and B.~Wen,
  ``Shadowdiffusion: When degradation prior meets diffusion model for shadow
  removal,'' in \emph{Proceedings of the IEEE/CVF Conference on Computer Vision
  and Pattern Recognition}, 2023, pp. 14\,049--14\,058.


\bibitem{kipf2016semi}
T.~N. Kipf and M.~Welling, ``Semi-supervised classification with graph
  convolutional networks,'' in \emph{International Conference on Learning
  Representations}, 2016.


\bibitem{kawahara2017brainnetcnn}
J.~Kawahara, C.~J. Brown, S.~P. Miller, B.~G. Booth, V.~Chau, R.~E. Grunau,
  J.~G. Zwicker, and G.~Hamarneh, ``Brainnetcnn: Convolutional neural networks
  for brain networks; towards predicting neurodevelopment,'' \emph{NeuroImage},
  vol. 146, pp. 1038--1049, 2017.




\bibitem{tzourio2002automated}
N.~Tzourio-Mazoyer, B.~Landeau, D.~Papathanassiou, F.~Crivello, O.~Etard,
  N.~Delcroix, B.~Mazoyer, and M.~Joliot, ``Automated anatomical labeling of
  activations in spm using a macroscopic anatomical parcellation of the mni mri
  single-subject brain,'' \emph{Neuroimage}, vol.~15, no.~1, pp. 273--289,
  2002.

\bibitem{wang2015gretna}
J.~Wang, X.~Wang, M.~Xia, X.~Liao, A.~Evans, and Y.~He, ``Gretna: a graph
  theoretical network analysis toolbox for imaging connectomics,''
  \emph{Frontiers in human neuroscience}, vol.~9, p. 386, 2015.

\bibitem{zuo2022constructing}
Q.~Zuo, L.~Lu, L.~Wang, J.~Zuo, and T.~Ouyang, ``Constructing brain functional
  network by adversarial temporal-spatial aligned transformer for early ad
  analysis,'' \emph{Frontiers in Neuroscience}, vol.~16, p. 1087176, 2022.

\bibitem{cui2013panda}
Z.~Cui, S.~Zhong, P.~Xu, Y.~He, and G.~Gong, ``Panda: a pipeline toolbox for
  analyzing brain diffusion images,'' \emph{Frontiers in human neuroscience},
  vol.~7, p.~42, 2013.



\bibitem{yu2020multi}
S.~Yu, S.~Wang, X.~Xiao, J.~Cao, G.~Yue, D.~Liu, T.~Wang, Y.~Xu, and B.~Lei,
  ``Multi-scale enhanced graph convolutional network for early mild cognitive
  impairment detection,'' in \emph{Medical Image Computing and Computer
  Assisted Intervention--MICCAI 2020: 23rd International Conference, Lima,
  Peru, October 4--8, 2020, Proceedings, Part VII 23}.\hskip 1em plus 0.5em
  minus 0.4em\relax Springer, 2020, pp. 228--237.

\bibitem{zhang2021deep}
L.~Zhang, L.~Wang, J.~Gao, S.~L. Risacher, J.~Yan, G.~Li, T.~Liu, D.~Zhu,
  A.~D.~N. Initiative \emph{et~al.}, ``Deep fusion of brain structure-function
  in mild cognitive impairment,'' \emph{Medical image analysis}, vol.~72, p.
  102082, 2021.

\bibitem{liu2022enhanced}
L.~Liu, Y.-P. Wang, Y.~Wang, P.~Zhang, and S.~Xiong, ``An enhanced multi-modal
  brain graph network for classifying neuropsychiatric disorders,''
  \emph{Medical Image Analysis}, vol.~81, p. 102550, 2022.

\bibitem{zuo2023brain}
Q.~Zuo, B.~Lei, N.~Zhong, Y.~Pan, and S.~Wang, ``Brain structure-function
  fusing representation learning using adversarial decomposed-vae for analyzing
  mci,'' \emph{IEEE Transactions on Neural Systems and Rehabilitation Engineering}, vol.~31, pp. 4017--4028, 2023.


\bibitem{lei2020self}
B.~Lei, N.~Cheng, A.~F. Frangi, E.-L. Tan, J.~Cao, P.~Yang, A.~Elazab, J.~Du,
  Y.~Xu, and T.~Wang, ``Self-calibrated brain network estimation and joint
  non-convex multi-task learning for identification of early alzheimer's
  disease,'' \emph{Medical image analysis}, vol.~61, p. 101652, 2020.

\bibitem{song2021graph}
X.~Song, F.~Zhou, A.~F. Frangi, J.~Cao, X.~Xiao, Y.~Lei, T.~Wang, and B.~Lei,
  ``Graph convolution network with similarity awareness and adaptive
  calibration for disease-induced deterioration prediction,'' \emph{Medical
  Image Analysis}, vol.~69, p. 101947, 2021.

\bibitem{qi2010impairment}
Z.~Qi, X.~Wu, Z.~Wang, N.~Zhang, H.~Dong, L.~Yao, and K.~Li, ``Impairment and
  compensation coexist in amnestic mci default mode network,''
  \emph{Neuroimage}, vol.~50, no.~1, pp. 48--55, 2010.

\bibitem{wang2021deficit}
X.~Wang, X.~Cui, C.~Ding, D.~Li, C.~Cheng, B.~Wang, and J.~Xiang, ``Deficit of
  cross-frequency integration in mild cognitive impairment and alzheimer's
  disease: A multilayer network approach,'' \emph{Journal of Magnetic Resonance
  Imaging}, vol.~53, no.~5, pp. 1387--1398, 2021.

\bibitem{chung2022association}
S.~J. Chung, Y.~J. Kim, J.~H. Jung, H.~S. Lee, B.~S. Ye, Y.~H. Sohn, Y.~Jeong,
  and P.~H. Lee, ``Association between white matter connectivity and early
  dementia in patients with parkinson disease,'' \emph{Neurology}, vol.~98,
  no.~18, pp. e1846--e1856, 2022.

\bibitem{lin2019multiparametric}
S.-Y. Lin, C.-P. Lin, T.-J. Hsieh, C.-F. Lin, S.-H. Chen, Y.-P. Chao, Y.-S.
  Chen, C.-C. Hsu, and L.-W. Kuo, ``Multiparametric graph theoretical analysis
  reveals altered structural and functional network topology in alzheimer's
  disease,'' \emph{NeuroImage: Clinical}, vol.~22, p. 101680, 2019.

\bibitem{bozzali2011anatomical}
M.~Bozzali, G.~J. Parker, L.~Serra, K.~Embleton, T.~Gili, R.~Perri,
  C.~Caltagirone, and M.~Cercignani, ``Anatomical connectivity mapping: a new
  tool to assess brain disconnection in alzheimer's disease,''
  \emph{Neuroimage}, vol.~54, no.~3, pp. 2045--2051, 2011.



\end{thebibliography}
\end{document}